%% file: acl_latex.tex
\documentclass[11pt]{article}

\usepackage[preprint]{acl}

\usepackage{times}
\usepackage{latexsym}

\usepackage[T1]{fontenc}

\usepackage[utf8]{inputenc}

\usepackage{microtype}

\usepackage{inconsolata}

\usepackage{microtype}
\usepackage{graphicx}
\usepackage{subfigure}
\usepackage{xcolor}
\usepackage[normalem]{ulem} 
\usepackage{booktabs}
\usepackage{amsfonts}
\usepackage{amsmath}
\usepackage{graphicx}
\usepackage{hyperref}
\usepackage{listings}
\usepackage{bbm}
\usepackage[most]{tcolorbox}

\lstdefinestyle{codestyle}{
    basicstyle=\ttfamily\footnotesize,
    breakatwhitespace=false,         
    breaklines=true,                 
    captionpos=b,                    
    keepspaces=true,                 
    showspaces=false,                
    showstringspaces=false,
    showtabs=false,                  
    tabsize=2
}
\newcommand{\name}[0]{SLMFix}

\title{\name{}: Leveraging Small Language Models for Domain Specific Language Error Fixing with Reinforcement Learning}

\author{
    David Jiahao Fu,
    Aryan Gupta,
    Aaron Councilman,
    Yu-Xiong Wang,
    Vikram Adve 
    \\
    University of Illinois Urbana-Champaign \\
    \texttt{\{jiahaof4, aryan10, aaronjc4, yxw, vadve\}@illinois.edu} \\
}

\begin{document}
\maketitle
\begin{abstract}
Large language models (LLMs) have shown impressive capabilities in code generation across many programming languages but even state-of-the-art LLMs generate programs that contain syntactic errors and fail to complete the given tasks, especially for low-resource programming languages (LRPLs). In addition, the high cost of training makes finetuning LLMs unaffordable for those with constrained computational resources, further weakening the effectiveness of LLMs for code generation. In this work, we propose \textbf{\name{}}, a novel code generation pipeline that leverages a small language model (SLM) finetuned using reinforcement learning (RL) techniques to fix syntactic errors in LLM-generated programs for domain-specific languages (DSLs) based on interpreter feedback. Our experimental results demonstrate the effectiveness and generalizability of our approach across multiple DSLs, improving the validator pass rates by 40\% on LRPLs and eliminating more than 50\% of syntactic errors for high-resource DSLs. Notably, \name{} brings substantial performance improvement to the base model and outperforms supervised finetuning approach even for 7B models on LRPLs including Ansible and Lean, showing the potential of our approach in improving the quality of LLM-generated programs. 
\end{abstract}

\section{Introduction}
\input{paragraphs/introduction}

\section{Related Work}
\input{paragraphs/related-work}

\section{Methodology}
\input{paragraphs/method}

\section{Evaluation}
\input{paragraphs/experiments}

\section{Conclusion}

\input{paragraphs/conclusion}

\section*{Limitations}

We acknowledge several limitations of our work. First, the semantic similarity metrics we designed are still not accurate enough to fully replace test suites. Although our analysis showed that semantic similarity measurements are suitable for DSLs based on their characteristics and show high correlation with other metrics including LLM judgment and pass rate on test suites, they are too strict and have low true positive rates. Second, we only conducted experiments on four DSLs, including two high-resource languages and two low-resource languages, which cannot completely validate the effectiveness of the proposed approach across all DSLs. Conducting experiments on other DSLs could help to strengthen our claims. Third, our method relies on an efficient syntactic checker for the target language. Although this is generally available for most languages, manually implementing such a checker might be needed for certain languages like Ansible, which limits the usability of our pipeline. 

\section*{Acknowledgments}

We gratefully acknowledge David Grove at IBM for the guidance and helpful discussions in this work, as well as the support from the IBM-ILLINOIS Discovery Accelerator Institute (IIDAI). Any opinions, findings, or conclusions expressed in this material are those of the authors and do not necessarily reflect the views of IBM.

This research used the Delta advanced computing and data resource which is supported by the National Science Foundation (award OAC 2005572) and the State of Illinois. Delta is a joint effort of the University of Illinois Urbana-Champaign and its National Center for Supercomputing Applications. This use was through allocation CIS250276 from the Advanced Cyberinfrastructure Coordination Ecosystem: Services \& Support (ACCESS) program, which is supported by U.S. National Science Foundation grants \#2138259, \#2138286, \#2138307, \#2137603, and \#2138296.

\bibliography{references}

\appendix

\input{paragraphs/appendix}

\end{document}

%% file: paragraphs/introduction.tex
Program synthesis, a process that automatically generates a program from some (hopefully higher-level) description, has attracted the attention of many researchers since almost the beginning of the modern study of computer science \cite{manna_towards_1971,waldinger_knowledge_1975,manna_deductive_1979}. As large language models (LLMs) have shown strong performance in complex tasks, researchers have explored the possibility of leveraging LLMs for code generation, and recent state-of-the-art models including GPT-5 \cite{openai_gpt-5_2025} and Gemini \cite{comanici_gemini_2025} have achieved performance comparable to top-tier human programmers \cite{li_competition-level_2022, el-kishky_competitive_2025}, while agentic systems like GPT Codex \cite{chen_evaluating_2021} and Claude \cite{anthropic_claude_nodate} have seen wide adoption in industry.

Despite their success in code generation tasks across many programming languages, the current training and inference pipeline that leverages supervised finetuning (SFT) techniques for direct code generation still faces several critical challenges. First, unlike a typical human programming approach, LLMs directly generate the output in the target language based on the input query without testing the output. This approach has no safeguards against incorrect code, and even state-of-the-art LLMs are prone to syntactic errors, semantic mistakes, and possible model hallucinations \cite {song_empirical_2023, zhang_llm_2025}. Second, to fully understand the syntax of the target language, LLMs require a large dataset for training, which does not exist for low-resource programming languages (LRPLs) and is often difficult to construct due to the scarcity and fragmentation of high-quality training data for the target language and limited community usage. As a result, LLMs tend to show inferior performance on LRPLs \cite{orlanski_measuring_2023}. Finally, achieving strong results generally requires finetuning the LLM, which is time-consuming and hardware-intensive, especially for very large models.

Prior works attempt to tackle these problems by proposing advanced training techniques \cite{gong_multicoder_2022,zhang_bridge-coder_2025} or pipelines involving program fixing modules based on compiler feedback \cite{zhang_self-edit_2023,wang_compilable_2022}. However, these methods only target one of the challenges and a unified pipeline that can effectively solve all three challenges is still missing. 

In this paper, we focus on addressing these challenges to improve the quality of LLM-generated programs for domain-specific languages (DSLs). While general-purpose programming languages have vast datasets and well-established tooling, DSLs are often low-resource and lack large-scale datasets, which makes training LLMs for reliable code generation particularly challenging. Specifically, our goal is to eliminate statically detected errors, such as syntax and type errors, in LLM-generated programs, which are easy to detect but remain challenging for existing LLMs to eliminate \cite{xia_automated_2023}. 

Recent studies \cite{kusama_how_2025, sheng_slm-sql_2025} have shown the potential of small language models (SLMs) for code-related tasks, and the task of fixing syntactic errors could be very suitable for SLMs because the model only needs to be able to make corrections, not generate the code from scratch. Meanwhile, Reinforcement learning (RL) approaches \cite{le_coderl_2022,liu_rltf_2023,jha_rlsf_2024} are widely employed in training LLMs to generate high-quality programs because they support more flexible training targets compared to SFT, but also bring higher training costs compared to SFT. However, using RL techniques to train SLMs would be much cheaper yet remaining effective.

Based on this finding, we propose \name{}, a novel code generation pipeline that can be applied to any existing LLMs pretrained on general code generation tasks without finetuning on the model, and performs well even for DSLs with limited training data. We achieve this by incorporating an SLM with 500M parameters finetuned with RL for error-fixing in the target language.

To ensure that SLM fixes the errors while aligning to the original prompt, we use a reward which combines a static validator to check both syntactic and other statically checkable correctness properties, and a semantic scorer that evaluates the functional correctness. The static validation can be performed by a variety of static tools, such as parsers, type-checkers, linters, and even symbolic interpreters, but semantic evaluation is challenging. Existing approaches \cite{zhang_self-edit_2023,le_coderl_2022,orlanski_measuring_2023} evaluate functional correctness of the generated program using sets of test cases but constructing such test suites requires significant effort and is necessarily far from complete since the range of valid programs and input data values for those programs is nearly unbounded. Instead, we build on prior works \cite{liang_grammar-based_2025, song_revisiting_2024} that showed comparing the similarities of the abstract syntax trees (ASTs) parsed from the generated program and the corresponding ground-truth program can be an effective method of checking functional correctness without test suites. This approach is especially well-suited to DSLs, whose task-specific and well-structured nature limits the number of programs that achieve the same functionality. With careful analysis (outlined in Appendix \ref{appendix:metrics}), we found that AST similarity is able to achieve more than 75\% accuracy compared to testing.

To address the issue of data scarcity, we construct the training dataset for the finetuning by first manually crafting a very small dataset containing a matching set of natural language query and ground truth program that have a good coverage of the target language. We then employ LLMs to generate corresponding (potentially incorrect) programs as examples for error-fixing, which can significantly increase the amount of training data available.

We evaluate our proposed approach on four example DSLs: Ansible, Bash, SQL, and Lean. While Bash and SQL are high-resource languages with large scale training data, Ansible and Lean, which are popular tools for IT automation and theorem proving, respectively, remain low-resource languages due to limited dataset availability. To verify the effectiveness of our method on Ansible, we constructed an Ansible dataset consisting of Ansible playbooks scraped from open-source Ansible repositories. With comprehensive experiments on multiple LLMs and baselines, we show that \name{} is able to bring substantial improvements to the base models across multiple DSLs while being very efficient, increasing the pass rate of static checker over 40\% for LRPLs and fixing more than 50\% of the static errors for high-resource DSLs, validating the effectiveness of our approach.

In summary, our contributions are as follows:
\begin{itemize}
    \item We propose \name{}, a novel code generation pipeline for DSLs that leverages a pretrained LLM for initial code generation and an SLM fine-tuned using RL for error-fixing, and show it can significantly improve the quality of LLM-generated programs with low latency across multiple DSLs, including both high-resource and low-resource languages, which demonstrates the benefits of using SLMs for error fixing. 
    \item We demonstrated the effectiveness of using semantic similarity based on ASTs to evaluate the functional correctness of LLM-generated programs in DSLs that can work as an alternative for a well-designed test suite.
    \item We constructed a new dataset for Ansible code generation, scraped from open-source repositories, that can be utilized for Ansible program generation training and evaluation. 
\end{itemize}

%% file: paragraphs/related-work.tex
\subsection{Reinforcement Learning for Code Gen}

Reinforcement Learning (RL) has seen wide adoption in modern LLM training because it supports dynamic measurements of model performance during training and more flexible training targets compared to SFT with an objective of next token prediction. For code generation, RL methods are able to consider the syntactic and functional correctness by leveraging compilers and test suites to evaluate the generated code beyond token matching, making them more suitable for training than SFT. CodeRL \cite{le_coderl_2022} proposed training a separate LLM to predict the functional correctness of generated programs using unit test results and the reference program. Subsequent works have improved this approach by introducing more fine-grained reward functions and training schemes \cite{shojaee_execution-based_2023, dou_stepcoder_2024, liu_rltf_2023, jha_rlsf_2024, ye_process-supervised_2025}. 

However, RL approaches generally require a test suite to verify the functional correctness of the code, which is expensive to construct and error-prone. To address this issue, recent works \cite{wang_compilable_2022, jain_tuning_2023} also proposed new training pipelines that learn from compiler feedback to generate more compilable codes without requiring test cases. 

\subsection{Automated Error Fixing}

Automated error fixing leverages LLMs to diagnose and fix errors in code and is commonly employed in the code generation pipelines to improve the quality of LLM-generated code. Prior works have explored training-free methods \cite{ngassom_chain_2024,shi_code_2024}, supervised finetuning \cite{zhang_self-edit_2023, liu_fastfixer_2024}, and RL-based methods \cite{gupta_deep_2019,zhao_repair_2024} for automated error fixing, showing that error fixing can significantly improve the quality of generated programs. However, most of these approaches require large scale training data, which limits their effectiveness in LRPLs. 

\subsection{Small Language Models}

Small Language Models (SLMs) have received lots of attention recently from the AI community as their smaller model size makes training and inference more efficient. Previous works \cite{sheng_slm-sql_2025,kusama_how_2025,koutcheme_benchmarking_2024} have explored training SLMs for code generation tasks and found that SLMs with 3B parameters can achieve comparable performance or even outperform state-of-the-art LLMs with over 10B parameters with proper training methods. \citet{kusama_how_2025} also attempted to employ SLMs for error fixing, but the authors only focused on high-resource languages like Python and Java which SLMs are already good at, while their capabilities on low-resource languages remain unexplored. 

%% file: paragraphs/method.tex
\begin{figure*}[!htb]
    \centering
    \resizebox{\textwidth}{!}{
        \includegraphics[width=0.5\linewidth]{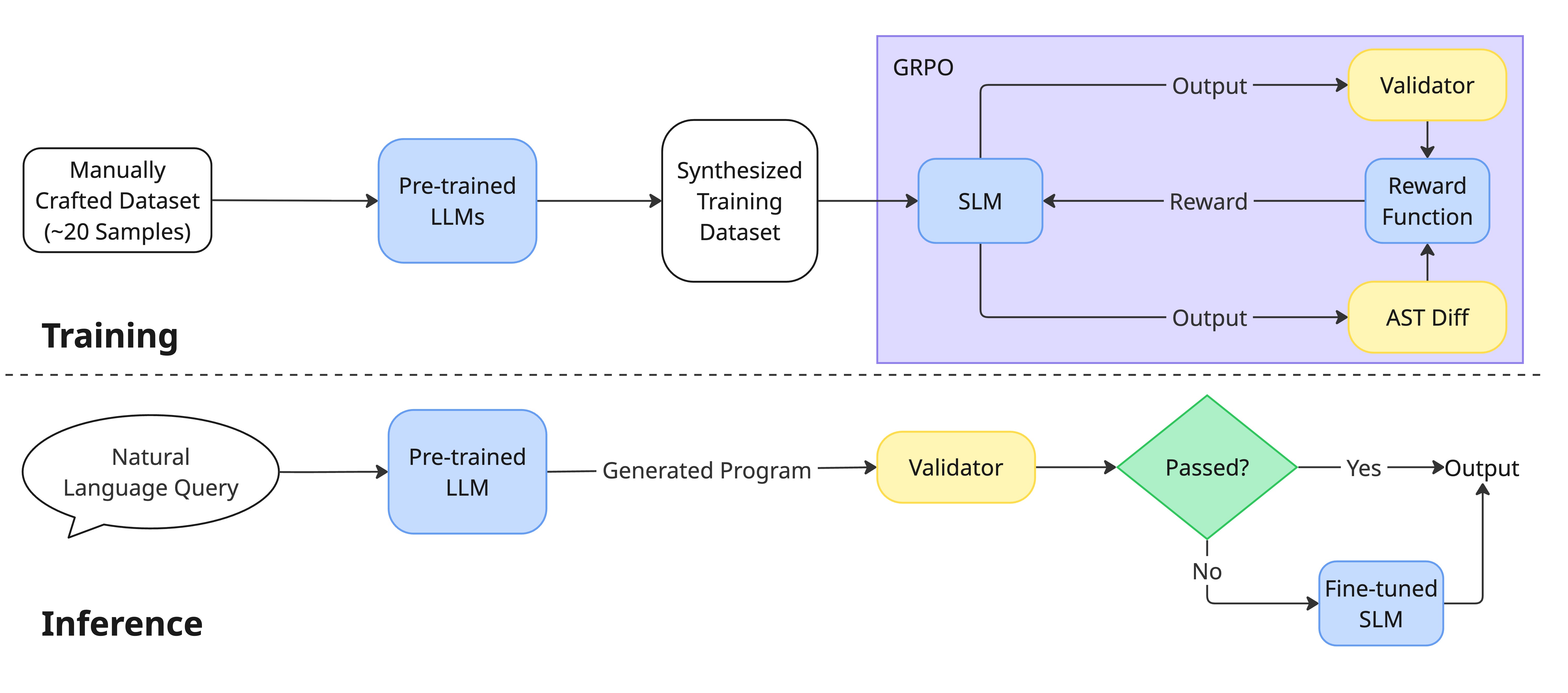}
    }
    \caption{The overview of the proposed training and inference pipelines.}
    \label{fig:ovewview}
\end{figure*}

Our proposed training and inference pipelines are summarized in Figure \ref{fig:ovewview}. We train a Small Language Model (SLM) with Group Relative Policy Optimization (GRPO) \cite{shao_deepseekmath_2024} to fix syntactic errors given the LLM-generated programs in a specific programming language and the feedback from a static validator of the language. During inference, our framework uses a frozen LLM pretrained for general code generation to generate an initial version of the program given the natural language query, which is then passed into the static validator to detect errors such as syntax and type errors. If there are no errors the program is output and otherwise the program, the original prompt, and the error messages from the validator are provided to the SLM which generates a revised program as the final output.

In the rest of this section we discuss our training method for the SLM (Section \ref{subsec:method-training}), the validator we employed in our training and evaluation (Section \ref{subsec:method-validator}), our approach to developing the semantic similarity metrics used in training (Section \ref{subsec:method-semantic}), and finally outline our process for constructing a new Ansible dataset (Section \ref{subsec:ansible-data}) and generating a large amount of training data from a small initial dataset (Section \ref{subsec:method-data}).

\subsection{Training}
\label{subsec:method-training}

Supervised finetuning with next token prediction as training objective does not reflect the focus on semantic correctness for code-related tasks that do not require token for token equivalence. Therefore, we use reinforcement learning with reward signals computed from the feedback of a static validator, which checks properties like syntax and types, and the score given by a semantic similarity metric, which compares the generated program with the ground truth to estimate the semantic correctness of the program. The static validator generates a boolean signal indicating if the given program contains syntactic errors, while the semantic metric will output a score between 0 and 1 representing how well the program aligns with the input query. The reward of the program is then a weighted average between these two signals, and the model parameters will be updated accordingly with Group Relative Policy Optimization (GRPO).

In order to balance between static and semantic correctness, we weight the scores from the validator and semantic metric in the reward function adaptively based on the percent of generated programs, in the current batch, that pass the validator. With this, the training focuses on improving static validity when most fixed programs still contain syntactic errors, and starts to focus on semantic correctness when fixed programs have a high pass rate on the validator. More formally, given a batch of programs $p_1, \dots, p_n$, the reward given to program $p_i$ would be 
$$r_i = (1-r_{pass})\cdot\mathbbm{1}[f_s(p_i)=true] + r_{pass} \cdot f_f(p_i)$$
where $f_s(p)\in \{true, false\}$ is the response of the static validator, $f_f(p)\in [0, 1]$ evaluates the semantic similarity between the program and the ground-truth, and $r_{pass} = \frac 1 n \sum \mathbbm{1}[f_s(p_i)=true]$ is the pass rate of the current batch. $r_{pass}$ is clamped at $0.5$ to avoid overemphasizing syntactic correctness.

\subsection{Validator Design}
\label{subsec:method-validator}

For validating Ansible playbooks, we use Astrogator \cite{councilman_towards_2025} which is a symbolic interpreter for Ansible. This system includes a parser for (a subset of) Ansible; performs type-checking of the Ansible program, including ensuring variables are well-defined and that module arguments are well typed; and ensures that arguments to module are appropriate (i.e., ensures that required arguments or mutually-exclusive arguments are provided appropriately). Additionally, it can perform symbolic interpretation of Ansible programs which simulates the behavior of the program on all different possible inputs but in this work, we only use the interpreter for static checking but the interpreter could also be used to run tests or to verify the behavior of the program matches some other reference, such as a formal specification~\cite{councilman_towards_2025}.

To validate Bash scripts, we use ShellCheck\footnote{https://github.com/koalaman/shellcheck} to identify syntactic errors. ShellCheck will also report checkstyle and linting errors, but we only consider syntax errors in this work. For SQL, we reconstructed the database according to the schema provided by the dataset and executed the generated queries to detect syntax errors. For Lean, we used the built-in parser to parse the program and find potential errors. For all validators, we only use the binary signal of whether the program contains syntactic errors because some validators might not provide detailed error messages. 

\subsection{Semantic Similarity Metric Design}
\label{subsec:method-semantic}

Domain-specific languages are typically task-specific, well-structured, and highly abstract languages. Therefore, commands and constructs in DSLs are generally high-level and tailored to a specific domain, which greatly limits the number of different ways to complete a specific task. From this observation, we propose to measure how well a generated program aligns with the input query by comparing the abstract syntax trees (ASTs) of the generated program and a reference ground-truth program. ASTs are the result of parsing the target language program and often abstract some syntactic details, meaning two syntactically different programs may have the same or very similar ASTs. The following subsections describe the language-specific AST-based semantic similarity metric we design for each language in our evaluation.

\subsubsection{Ansible}

Ansible programs (called playbooks), which we focus on generating, are composed of a sequence of \emph{plays} that define the operations to perform. Each play consists of a sequence of \emph{tasks} that define the atomic operations in the play, as well as other configurations that specify how the tasks should be executed. 

Motivated by \citet{pujar_invited_2023}, we decompose each playbook into a sequence of plays and divide each play into a list of tasks, and parse the arguments to each task into a dictionary. For each task in the ground-truth, we find the corresponding task in the generated playbook and check if each key in the task dictionary appears in the corresponding task dictionary of the generated task and if the same value is assigned to the key. The final score assigned to a particular task is 
$$score = \frac{\sum_{(k, v)\in \hat{y}} \mathbbm{1} [(k, v) \in y]}{|\hat{y}|}$$
where $\hat{y}$ is the task dictionary of ground-truth and $y$ is the task dictionary of the generated playbook. We then obtain the final score by averaging over all tasks in the play and all plays in the playbook.

\subsubsection{Bash}

Since there are no existing datasets for multi-line Bash script generation, we focus on single-line Bash scripts which can contain parentheses and connectors. We use Bashlex\footnote{https://github.com/idank/bashlex} to parse a Bash script into \emph{atomic commands}, where each atomic command executes a single bash command without any connectors. Similar to Ansible, we then parse each atomic command into a dictionary, from the flags and value passed to a command, and check if each flag in the ground-truth appears in the output and is associated with the same value. We obtain the overall score for a Bash script by taking the average over all of the atomic commands.

\subsubsection{SQL}

For SQL, we use SQLGlot\footnote{https://github.com/tobymao/sqlglot} to normalize a query and parse it into an AST. To improve the accuracy of the AST-based comparison, we remove all column aliases and replace table aliases with their original names. The semantic similarity score is then computed as the reciprocal of the tree edit distance between the AST of the generated program and the AST of the ground-truth. 

\subsubsection{Lean}

To compare generated Lean programs against ground-truth, we examine from three aspects: overall token overlap, declaration overlap, and tactic overlap, which capture the most important parts of a Lean program. We calculate the matching in each aspect with the F-1 score and compute the semantic similarity with the average of these scores. 

\subsection{Ansible Dataset Construction}
\label{subsec:ansible-data}

Since no existing dataset is publicly available for Ansible, we constructed an Ansible dataset through an end-to-end collection and parsing pipeline to gather real-world Ansible code. The data collection process began with the development of a custom web scraper built using Selenium to systematically crawl Ansible Galaxy\footnote{https://galaxy.ansible.com/}, the primary community hub for Ansible that contains Ansible repositories with open-source licenses, to identify and download all publicly available repositories listed on the platform.

Following the repository collection phase, the downloaded content underwent structured parsing using the Ansible Content Parser\footnote{https://github.com/ansible/ansible-content-parser}. This specialized parser analyzes Ansible files to extract valid, parsable playbooks from the collected repositories and remove sensitive information from the repositories. The Ansible Content Parser examines the structure of Ansible content including playbooks, roles, and tasks, identifying syntactically correct playbooks while filtering out malformed or unparsable content. To ensure the quality of the playbooks, we performed additional filtering on the extracted playbooks, including deduplication and removal of playbooks with undefined variables or without any meaningful content.

A critical component of the parsing process involved variable and role infilling; these steps produce stand-alone executable programs rather than programs that rely on other files to resolve variables and imported roles. For variable infilling, we used the parser to identify variable references within playbooks and populated them with appropriate values based on variable definitions found in defaults, vars files, inventory configurations, and other sources following Ansible's variable precedence rules. For role infilling, we collected all roles that are defined in the scope of the playbook and replaced the roles by their corresponding definitions in the playbook.

The web scraping and parsing pipeline successfully yielded a corpus of 616 syntactically valid Ansible playbooks but these playbooks still lacked accompanying natural language descriptions as used as the inputs for the LLMs to generate the playbook. Therefore, we employed GPT-4o mini \cite{openai_gpt-4_2023} to generate natural language descriptions for the collected playbooks. To ensure the correctness of LLM labeled data, we conducted manual inspection on the generated descriptions, and details of the dataset and example inputs/outputs can be found in Appendix \ref{appendix:dataset}.

\subsection{Training Data Synthesis}
\label{subsec:method-data}

To construct training data for our reinforcement learning, we use a multi-model generation strategy that leverages the diversity of outputs from different large language models to improve robustness and reduce overfitting. From each language-specific dataset, we manually constructed a subset of 20 diverse natural language queries along with corresponding ground truth code implementations as the example set. With this small set of curated examples, we generate 10 sample programs for each example from five strong code LLMs: StarCoder-2 7B \cite{lozhkov_starcoder_2024}, DeepSeek-Coder 6.7B \cite{guo_deepseek-coder_2024}, Qwen2.5-Coder 7B \cite{hui_qwen25-coder_2024}, Granite 8B Code \cite{mishra_granite_2024}, and LLaMA-3.1 8B \cite{dubey_llama3_2024}.

This multi-model generation process creates an expansive training dataset 50 times the size of the original dataset consisting of natural language queries paired with both LLM-generated code (of varying quality) and corresponding ground truth implementations. The resulting dataset provides the varied examples necessary for training a strong and robust error-fixing model. 

%% file: paragraphs/experiments.tex
\begin{table*}[!htb]
    \centering
    \resizebox{\textwidth}{!}{
    \begin{tabular}{l|ccc|ccc|ccc|ccc|}
        \toprule
        & \multicolumn{3}{|c|}{Ansible} & \multicolumn{3}{|c|}{Bash} & \multicolumn{3}{|c|}{SQL} & \multicolumn{3}{|c|}{Lean}\\
        & CodeBERTScore & Pass Rate & AST Diff & CodeBERTScore & Pass Rate & AST Diff & CodeBERTScore & Pass Rate & AST Diff & CodeBERTScore & Pass Rate & AST Diff \\ 
        \midrule
        \textbf{GPT-5.4 mini} & & & & & & & & & & & & \\
        \quad Base & 0.8010 & 25.00\% & 0.3240 & 0.8688 & 99.66\% & 0.5062 & 0.8905 & 99.32\% & 0.5383 & 0.7535 & 84.42\% & 0.1189 \\
        \quad ICL & \textbf{\underline{0.8421}} & 71.31\% & \underline{0.3573} & \underline{0.8777} & 99.66\% & \underline{0.5408} & \underline{0.9250} & 99.32\% & \underline{0.5909} & \underline{0.7704} & 90.00\% & \textbf{\underline{0.1287}} \\ 
        \quad Self-correction & 0.8015 & 36.74\% & 0.2839 & 0.8688 & \textbf{\underline{100.00\%}} & 0.5075 & 0.8707 & \textbf{\underline{99.90\%}} & 0.5398 & 0.7552 & \textbf{\underline{93.85\%}} & 0.1220 \\ 
        \quad Self-Edit & 0.7508 & 59.23\% & 0.2064 & 0.7525 & \textbf{\underline{100.00\%}} & 0.5054 & 0.8373 & 99.61\% & 0.5380 & 0.7296 & 85.19\% & 0.1088 \\
        \quad \name{} (\textbf{Ours}) & 0.8346 & \underline{92.50\%} & 0.3066 & 0.8487 & 99.77\% & 0.5045 & 0.8800 & 99.74\% & 0.5384 & 0.7454 & 89.91\% & 0.1182\\
        & \small (0.0035) & \small (0.0063) & \small (0.0016) & \small (0.0031) & \small (0.0008) & \small (0.0000) & \small (0.0137) & \small (0.0005) & \small (0.0003) & \small (0.0002) & (0.0002) & \small (0.0000) \\

        \midrule
        \textbf{Qwen3-Coder 30B A3B} & & & & & & & & & & & & \\
        \quad Base & 0.8121 & 44.63\% & 0.2961 & 0.869 & 99.66\% & 0.4883 & 0.8837 & 99.61\% & 0.5929 & 0.7222 & 39.23\% & 0.0778 \\ 
        \quad ICL & \underline{0.8394} & 61.74\% & \underline{0.3346} & \underline{0.8783} & 99.49\% & \underline{0.5136} & \underline{0.9211} & 98.74\% & \underline{0.6267} & \underline{0.7795} & 46.73\% & \underline{0.1010} \\
        \quad Self-correction & 0.8119 & 53.36\% & 0.3086 & 0.8692 & 99.83\% & 0.4883 & 0.8838 & \textbf{\underline{99.90\%}} & 0.5933 & 0.7341 & 44.62\% & 0.0803 \\
        \quad Self-Edit & 0.7452 & 72.32\% & 0.2706 & 0.7540 & \textbf{\underline{100.00\%}} & 0.4883 & 0.8388 & 99.61\% & 0.5929 & 0.7117 & 46.54\% & 0.0548 \\
        \quad \name{} (\textbf{Ours}) & 0.8390 & \underline{94.02\%} & 0.3218 & 0.8493 & 99.83\% & 0.4883 & 0.8824 & 99.68\% & 0.5930 & 0.6640 & \underline{79.81\%} & 0.0604 \\
        & \small (0.0007) & \small (0.0348) & \small (0.0032) & \small (0.0035) & \small (0.0000) & \small (0.0000) & \small (0.0133) & \small (0.0005) & \small (0.0000) & \small (0.0001) & \small (0.0043) & \small (0.0001) \\

        \midrule
        \textbf{Qwen-2.5-Coder 7B} & & & & & & & & & & & & \\
        \quad Base & 0.8198 & 59.56\% & 0.3097 & 0.8692 & 98.46\% & 0.4948 & 0.9082 & 95.74\% & 0.5650 & 0.7320 & 15.96\% & 0.0881 \\
        \quad SFT & 0.8392 & 73.83\% & \textbf{\underline{0.3704}} & \textbf{\underline{0.8923}} & 99.32\% & \textbf{\underline{0.5566}} & \textbf{\underline{0.9531}} & 94.29\% & \textbf{\underline{0.6473}} & \underline{0.7878} & 7.31\% & 0.0988 \\
        \quad ICL & 0.8385 & 65.94\% & 0.3455 & 0.8751 & \underline{99.83\%} & 0.5152 & 0.9418 & 96.71\% & 0.6318 & 0.7734 & 54.23\% & \underline{0.1001} \\
        \quad Self-correction & 0.8205 & 60.74\% & 0.3136 & 0.8693 & 98.63\% & 0.4961 & 0.9072 & 97.20\% & 0.5657 & 0.7340 & 16.54\% & 0.0885 \\
        \quad Self-Edit & 0.7399 & 75.84\% & 0.2958 & 0.7502 & \underline{99.83\%} & 0.4987 & 0.8363 & 96.52\% & 0.5648 & 0.7253 & 23.27\% & 0.0503 \\
        \quad \name{} (\textbf{Ours}) & \underline{0.8398} & \textbf{\underline{96.92\%}} & 0.3382 & 0.8492 & \underline{99.83\%} & 0.4984 & 0.8830 & \underline{97.45\%} & 0.5682 & 0.6598 & \underline{75.48\%} & 0.0640 \\
        & \small (0.0007) & \small (0.0174) & \small (0.0054) & \small (0.0029) & \small (0.0000) & \small (0.0001) & \small (0.0119) & \small (0.0030) & \small (0.0002) & \small (0.0002) & \small (0.0023) & \small (0.0000) \\

        \midrule
        \textbf{DeepSeek-Coder 6.7B} & & & & & & & & & & & & \\
        \quad Base & 0.8203 & 59.40\% & 0.3319 & 0.8511 & \textbf{\underline{100.00\%}} & 0.4359 & 0.8780 & 98.65\% & 0.4609 & 0.7292 & 9.62\% & 0.0977 \\
        \quad SFT & 0.8087 & 47.54\% & 0.3081 & \underline{0.8794} & 97.95\% & \underline{0.5240} & \underline{0.9189} & 94.78\% & \underline{0.5753} & \underline{0.7552} & 21.35\% & 0.0890 \\
        \quad ICL & 0.8309 & 61.91\% & \underline{0.3530} & 0.8603 & 99.49\% & 0.4652 & 0.8965 & 97.49\% & 0.5145 & 0.7469 & \underline{51.73\%} & 0.0100 \\
        \quad Self-correction & 0.8178 & 62.75\% & 0.3336 & 0.8502 & \textbf{\underline{100.00\%}} & 0.4359 & 0.8778 & \underline{99.81\%} & 0.4619 & 0.7331 & 10.00\% & \underline{0.1007} \\
        \quad Self-Edit & 0.7395 & 77.52\% & 0.2989 & 0.7541 & \textbf{\underline{100.00\%}} & 0.4359 & 0.8376 & 98.94\% & 0.4607 & 0.7283 & 16.35\% & 0.0562 \\
        \quad \name{} (\textbf{Ours}) & \underline{0.8368} & \underline{95.80\%} & 0.3408 & 0.8445 & \textbf{\underline{100.00\%}} & 0.4359 & 0.8806 & 99.29\% & 0.4639 & 0.6754 & 42.02\% & 0.0714 \\
        & \small (0.0002) & \small (0.0261) & \small (0.0014) & \small (0.0018) & \small (0.0000) & \small (0.0000) & \small (0.0118) & \small (0.0004) & \small (0.0001) & \small (0.0003) & \small (0.0000) & \small (0.0000) \\

        \midrule
        \textbf{Granite-3.3 8B} & & & & & & & & & & & & \\
        \quad Base & 0.8147 & 45.30\% & 0.2891 & 0.8412 & 98.63\% & 0.4143 & 0.8759 & 90.23\% & 0.3929 & 0.7585 & 33.65\% & 0.1101 \\
        \quad SFT & 0.8298 & 82.89\% & 0.2945 & 0.8342 & 93.16\% & 0.3850 & \underline{0.9074} & 64.89\% & 0.3820 & \textbf{\underline{0.7888}} & 10.77\% & 0.1012 \\
        \quad ICL & 0.8331 & 54.53\% & \underline{0.3262} & \underline{0.8546} & 98.97\% & \underline{0.4520} & 0.8831 & 91.39\% & \underline{0.4154} & 0.7859 & 46.15\% & \underline{0.1164} \\
        \quad Self-correction & 0.8134 & 51.68\% & 0.3014 & 0.8402 & 98.97\% & 0.4143 & 0.8715 & \underline{95.65\%} & 0.3987 & 0.7597 & 37.88\% & 0.1110 \\
        \quad Self-Edit & 0.7405 & 66.28\% & 0.2356 & 0.7516 & 99.32\% & 0.4169 & 0.8331 & 91.59\% & 0.3928 & 0.7280 & 39.04\% & 0.0732 \\
        \quad \name{} (\textbf{Ours}) & \underline{0.8408} & \underline{95.30\%} & 0.3105 & 0.8428 & \underline{99.38\%} & 0.4159 & 0.8796 & 93.87\% & 0.4028 & 0.6740 & \underline{80.29\%} & 0.0817 \\
        & \small (0.0007) & \small (0.0142) & \small (0.0006) & \small (0.0018) & \small (0.0008) & \small (0.0005) & \small (0.0114) & \small (0.0057) & \small (0.0011) & \small (0.0010) & \small (0.0083) & \small (0.0000) \\

        \bottomrule
    \end{tabular}
        }
    \caption{Evaluation results on Ansible, Bash, SQL, and Lean code generation. The best performance in each column is marked in \textbf{bold}, while the best performance in each group is marked in \underline{underline}. For \name{}, we report the average and variance over three runs, with the variance shown in parentheses under the average. \name{} is able to achieve the highest pass rate on the validator across all languages and LLMs, achieving competitive performance and even outperforms the strongest baselines on semantic metrics.}
    \label{tab:llm-result}
\end{table*}

\subsection{Experimental Setup}

\textbf{Target Programming Languages.} We chose Ansible, Bash, SQL, and Lean as example DSLs for our evaluation for a diversity of languages, including high- and low-resource languages. Bash and SQL are high-resource DSLs with abundant training data, while Ansible and Lean are examples of low-resource programming languages despite being widely used in many fields.

\textbf{Datasets.} For Ansible code generation, we utilized the dataset we constructed, as described in Section \ref{subsec:ansible-data}. For Bash we used the NL2Bash dataset \cite{lin_nl2bash_2018}, a corpus that covers over 100 commonly used Bash utilities with diverse commands. For SQL command generation, we leveraged the Spider dataset \cite{yu_spider_2018}, a large-scale human-labeled corpus designed specifically for complex and cross-domain semantic parsing and text-to-SQL tasks. The natural language queries were augmented with their corresponding database schemas to provide models with the necessary context to understand the database structure and generate correct SQL queries that reference valid table and column names. For Lean, we employed the NuminaMath \cite{wang_kimina-prover_2025} dataset, which consists of mathematical competition problems formalized in Lean 4. For all datasets, we removed the samples that cannot pass our validator. Details about the dataset split and leakage prevention can be found in Appendix \ref{appendix:training}.

\textbf{Baselines.} To evaluate the performance of our proposed pipeline for both in-distribution and out-of-distribution data, we selected a variety of strong coding LLMs, including GPT-5.4 mini \cite{openai_gpt-5_2025}, Qwen3-Coder 30B \cite{team_qwen3_2025}, Qwen-2.5-Coder 7B \cite{hui_qwen25-coder_2024}, DeepSeek-Coder 6.7B \cite{guo_deepseek-coder_2024}, and Granite-3.3 8B \cite{granite2024granite} models as the base LLMs, and compared our proposed approach against the following types of baselines in our experiments:
\begin{itemize}
    \item Pre-trained LLMs: directly generate the program given the input query;
    \item Supervised fine-tuning (SFT): finetune the base LLM on target language with the training dataset and employ finetuned LLMs for direct code generation;
    \item In-context learning (ICL): provide a set of five samples from the curated example set of the target language to the base LLM for direct code generation;
    \item Self-correction: provide the same LLM with the error message from the validator and ask it to revise the original program it generated;
    \item Automatic program repair methods: finetune the SLM for error-fixing with Self-Edit \cite{zhang_self-edit_2023}, the example of automatic program repair methods we selected.
\end{itemize}

Due to consideration of finetuning cost, we only performed SFT on Qwen-2.5-Coder 7B, DeepSeek-Coder 6.7B, and Granite-3.3 8B. For our error-fixing model, we used Qwen-2.5 Coder 0.5B \cite{hui_qwen25-coder_2024} which we fine-tune for the task. For in-context learning, we randomly selected five samples from the training dataset as the in-context examples.

\textbf{Metrics.} For all languages, we report CodeBERTScore \cite{zhou_codebertscore_2023} to evaluate the semantic similarity between the generated program and the ground-truth. For each language, we also report the pass rate on the corresponding validator and the average score on the semantic similarity metric described in Section \ref{subsec:method-semantic}.

\subsection{Implementation Details}

For reinforcement learning approaches, we use verl \cite{sheng2024hybridflow} and vLLM \cite{kwon_vllm_2023} to perform training with 70 total epochs and batch size of 64 on the Qwen-2.5-Coder 0.5B model. We set the maximum input and output length to be 4096 tokens, and the learning rate for the model as 1e-6, respectively. For supervised finetuning, we employed LoRA \cite{hu_lora_2022} for parameter-efficient finetuning. We set the maximum number of steps to be 150 and batch size as 4, with learning rate of 1e-4 and weight decay of 0.05. At inference time, we use vLLM for efficient generation, setting temperature to 0 to avoid randomness. For \name{}, we report the average and variance over three training runs. We conducted all experiments on two 40GB A40 GPUs and the experiments took roughly 100 GPU hours in total to complete. Details about the RL feedback can be found in Appendix \ref{appendix:training}.

\subsection{Main Results}

Table \ref{tab:llm-result} presents the evaluation results of \name{} and baselines on Ansible, Bash, SQL, and Lean code generation using different LLMs as the base model. \name{} is able to significantly improve the syntactic correctness of LLM-generated programs while remaining aligned with the user intent. For Ansible and Lean, which are low-resource languages, \name{} is able to improve the pass rate of static checking by an average of over 40\%, which is higher than most baselines we compared against. For Bash and SQL, although the base LLMs are already able to eliminate most syntactic errors in the generated programs, \name{} is still able to further improve the quality of the programs by correcting more than 50\% of the static errors in the output. Results on CodeBERTScore and AST Diff also show that \name{} is able to preserve the semantic meaning of the original program while fixing syntactic errors, verifying the effectiveness of our proposed pipeline.

To further strengthen the results, we additionally conducted evaluations with the SQL test suites \cite{ruiqi20}, which is outlined in Table \ref{tab:test-sql}. The results show that the AST Diff metric aligns well with the testing results and validate AST Diff as a useful metric to measure the functional correctness of the programs.

\begin{table}[!htb]
    \centering
    \resizebox{\columnwidth}{!}{
    \begin{tabular}{l|cccccc|}
        \toprule
         & Base & SFT & ICL & Self-correction & Self-Edit & SLMFix \\
         \midrule
         GPT-5.4 mini & 0.676 & - & \textbf{0.706} & 0.677 & 0.676 & 0.676 \\
         Qwen3-Coder 30B A3B & 0.709 & - & \textbf{0.725} & 0.709 & 0.709 & 0.709 \\
         Qwen-2.5-Coder 7B & 0.699 & \textbf{0.716} & \textbf{0.716} & 0.704 & 0.699 & 0.702 \\
         DeepSeek-Coder 6.7B & 0.587 & \textbf{0.652} & 0.608 & 0.590 & 0.587 & 0.590 \\
         Granite-3.3 8B & 0.531 & 0.376 & 0.538 & \textbf{0.551} & 0.533 & 0.544 \\
        \bottomrule
    \end{tabular}
    }
    \caption{Comparison of baselines and SLMFix on SQL test suites, where the best performance in each row is marked in \textbf{bold}. }
    \label{tab:test-sql}
\end{table}

\subsection{Efficiency Analysis}

To better understand the efficiency of \name{} compared to the baseline, we also analyzed the latency of \name{} in code generation, as shown in Table \ref{tab:efficiency}. The results show that \name{} takes almost no additional time after initial generation to fix syntactic errors, faster than most baselines across different LLMs. This result verifies that \name{} can be a very efficient code generation pipeline that can eliminate most syntactic errors. 

\begin{table}[!htb]
    \centering
    \resizebox{\columnwidth}{!}{
    \begin{tabular}{l|cccccc|}
        \toprule
        & Base & SFT & ICL & Self-correction & Self-Edit & \name{} (\textbf{Ours}) \\
        \midrule
        Qwen-3-Coder 30B A3B & 60.29 & - & 51.41 & 133.23 & 184.87 & 68.50 \\
        Qwen-2.5-Coder 7B & 42.52 & 132.84 & 75.58 & 94.75 & 172.34 & 45.73 \\
        DeepSeek-Coder 6.7B & 124.00 & 263.27 & 1009.67 & 370.35 & 255.20 & 126.72 \\
        Granite-3.3 8B & 278.67 & 179.31 & 429.52 & 587.36 & 406.83 & 282.06 \\
        \bottomrule
    \end{tabular}
    }
    \caption{Latency comparison of baselines and \name{} on Ansible code generation. Latency is measured as total generation time in seconds for the evaluation dataset of size 596. For program repair methods, the latency is calculated as \textbf{sum of generation (Base) and fixing}. \name{} brings little additional inference latency (7.35 milliseconds per sample on average).}
    \label{tab:efficiency}
\end{table}

\subsection{Ablation Studies}

To better understand how different components of the training approach affect the overall performance of our proposed framework, we conducted a series of ablation studies on the components. To highlight the performance difference across various settings, we select Ansible as the target language and employ Qwen-2.5 0.5B \cite{yang_qwen25_2024} and LLaMA-3.2 1B \cite{dubey_llama3_2024} as the base models to amplify the differences between design choices. Table \ref{tab:ablation-result} summarizes the results of the ablation study. We found that using only the static validator score or the semantic similarity metric would achieve excellent performance on the particular metric the model was trained on, but also cause the trained model to overfit to the training target while ignoring other metrics, resulting in serious performance degradation. In contrast, combining both metrics achieves a better balance between static validation and functional correctness, helping the model to generate higher quality programs. In addition, we also compared with using the proposed RL method to directly train the SLM on the code generation task to verify if using SLM for error fixing is needed. We found that the reinforcement learning method we utilized does not work well on direct code generation task, possibly due to the limited model capability of SLMs. This result demonstrates the benefits of leveraging SLMs for program repair over direct generation.

\begin{table}[!htb]
    \centering
    \tiny
    \resizebox{\columnwidth}{!}{
    \begin{tabular}{l|cccc|}
        \toprule
        & \multicolumn{4}{c|}{Ansible} \\
        & BLEU & CodeBERTScore & Pass Rate & AST Diff \\ 
        \midrule
        RL for code generation & 0.3856 & 0.8101 & 48.83\% & 0.2792 \\
        \midrule
        \textbf{Qwen-2.5 0.5B} & & & & \\
        \quad \name & \underline{0.3913} & 0.8274 & 94.30\% & 0.3455 \\
        \quad Syntactic score only & 0.3265 & \underline{0.8298} & \underline{98.99\%} & 0.2062 \\
        \quad Semantic score only & 0.2553 & 0.7588 & 22.48\% & \textbf{\underline{0.4047}} \\
        \midrule
        \textbf{LLaMA-3.2 1B} & & & & \\
        \quad \name & \textbf{\underline{0.4104}} & 0.8290 & 92.79\% & 0.3558 \\
        \quad Syntactic score only & 0.3612 & \textbf{\underline{0.8365}} & \textbf{\underline{99.50\%}} & 0.2353 \\
        \quad Semantic score only & 0.2462 & 0.7430 & 19.63\% & \underline{0.3601} \\
        \bottomrule
    \end{tabular}
    }
    \caption{The results of ablation studies. The best performance in each column is marked in \textbf{bold}, while the best performance in each group is marked in \underline{underline}. Our proposed training method can achieve an excellent balance between the syntactic and functional correctness compared to the ablated settings.}
    \label{tab:ablation-result}
\end{table}

%% file: paragraphs/conclusion.tex
In this paper, we present \name{}, a new pipeline for code generation that involves a pretrained LLM and an SLM tailored for error-fixing. We propose to leverage deep reinforcement learning techniques to train the small language model to fix identified errors in LLM-generated programs, using a fine-grained reward function that evaluates both syntactic and semantic correctness of the fixed programs. This approach is especially well suited to DSLs as they typically have well-structured syntax and high-level functions. We also verified the effectiveness of using similarity of abstract syntax trees as an estimate of functional correctness for DSLs to replace the need for a well-designed test suite, making training for code generation tasks much easier. With extensive experiments on Ansible, Bash, SQL, and Lean, we demonstrated that our proposed method is able to significantly improve the quality of LLM-generated programs by reducing syntactic errors, outperforming similar program repair approaches and achieving comparable performance with LLM finetuned on the task.

%% file: paragraphs/appendix.tex
\section{Details of Ansible Dataset}

\label{appendix:dataset}

Table \ref{tab:ansible_stats} shows the statistics of the data collection process. We only keep the playbooks that can be successfully parsed and validated, without any undefined variables or duplicate contents. 

\begin{table}[!htb]
    \centering
    \tiny
    \resizebox{\columnwidth}{!}{
        \begin{tabular}{l|cc|}
            \toprule
            \textbf{Category} & \textbf{Count} \\
            \midrule
            Repositories Scraped from Old Ansible Galaxy & 34057 \\
            Repositories Scraped from New Ansible Galaxy & 3212 \\
            Total Repositories Scraped & 37269 \\
            Total Playbooks Extracted & 74497 \\
            Playbooks with Undefined Variables & 5906 \\
            Playbooks with Duplicate Contents & 35033 \\
            Playbooks Successfully Parsed & 33558 \\
            Final Dataset Size & 616 \\
            \bottomrule
        \end{tabular}
    }
    \caption{Summary of Ansible repository and playbook collection statistics.}
    \label{tab:ansible_stats}
\end{table}

The following are some examples from the dataset. With manual inspections we found that the LLM-generated prompts align well with human-written prompts.

\begin{tcolorbox}[title={Example 1}]
Prompt: Generate an Ansible playbook that installs and configures a database on the localhost with root access.
\tcblower
Playbook: 
\begin{lstlisting}
---
- hosts: localhost
  remote_user: root
  tasks:
  - name: install postgresql
    yum: name=postgresql state=latest
    become: true
    become_method: sudo
\end{lstlisting}
\end{tcolorbox}

\begin{tcolorbox}[title={Example 2}]
Prompt: Generate an Ansible playbook that updates the apt package cache on Ubuntu hosts with elevated privileges.
\tcblower
Playbook: 
\begin{lstlisting}
---
- name: Prepare
  hosts: ubuntu
  become: true
  tasks:
  - name: Update apt packages
    apt:
      update_cache: true
\end{lstlisting}
\end{tcolorbox}

\begin{tcolorbox}[title={Example 3}]
Prompt: Generate an Ansible playbook that configures passwordless sudo access for the 'admin' group on macOS.
\tcblower
Playbook: 
\begin{lstlisting}
---
- name: Prepare
  hosts: macos
  become: true
  tasks:
  - name: Allow 'admin' group to have passwordless sudo
    ansible.builtin.lineinfile:
      dest: /etc/sudoers
      state: present
      regexp: ^%admin
      line: '%admin ALL=(ALL) NOPASSWD: ALL'
      validate: visudo -cf %s
\end{lstlisting}
\end{tcolorbox}

\section{Empirical Analysis of Semantic Similarity Metrics}
\label{appendix:metrics}

To verify the effectiveness of our proposed semantic similarity metrics, we compared the scores given by semantic similarity with the testing results on Ansible and SQL, two languages with existing test suites available.

For SQL, we utilized the evaluation set of the Spider dataset \cite{yu_spider_2018} with test suites provided by \citet{ruiqi20}. We randomly selected 100 outputs for each LLM we evaluated to run the test suites, and the comparison between execution match and semantic similarity is shown in Table \ref{tab:sql-result}. When considering having a semantic similarity of at least 0.8 as equivalent to the ground-truth, the metric is able to classify most outputs correctly with an accuracy of 76\%, a true positive rate of 64.2\% and a true negative rate of 98.8\%, which validates its effectiveness in estimating functional correctness.

\begin{table}[!htb]
    \centering
    \tiny
    \resizebox{\columnwidth}{!}{
        \begin{tabular}{l|cc|}
            \toprule
            & Tests Failed & Tests Passed \\
            \midrule
            $0.8\leq \text{Sim} \leq 1$ & 2 & 212 \\
            $0.6\leq \text{Sim} < 0.8$ & 0 & 0 \\
            $0.4\leq \text{Sim} < 0.6$ & 15 & 6 \\
            $0.2\leq \text{Sim} < 0.4$ & 24 & 35 \\
            $0\leq \text{Sim} < 0.2$ & 129 & 77 \\
            \bottomrule
        \end{tabular}
    }
    \caption{Comparison between execution match and semantic similarity for SQL. The overall accuracy of semantic similarity with respect to execution match is 76\%.}
    \label{tab:sql-result}
\end{table}

For Ansible, we employed the benchmark suite constructed by \citet{councilman_towards_2025}, which contains 21 questions and 1,260 outputs, and the comparison between testing result and semantic similarity is shown in Table \ref{tab:ansible-result}.  When considering having a semantic similarity of at least 0.8 as equivalent to the ground-truth, the metric is able to classify most outputs correctly with an accuracy of 84.2\% and a true negative rate of 94.6\%. However, the true positive rate is only 38.4\%, which indicates that the AST-based semantic similarity is too strict and fails to correctly classify Ansible programs that are functionally equivalent to the ground-truth but not semantically equivalent. 

\begin{table}[!htb]
    \centering
    \tiny
    \resizebox{\columnwidth}{!}{
        \begin{tabular}{l|cc|}
            \toprule
            & Tests Failed & Tests Passed \\
            \midrule
            $0.8\leq \text{Sim} \leq 1$ & 56 & 89 \\
            $0.6\leq \text{Sim} < 0.8$ & 89 & 89 \\
            $0.4\leq \text{Sim} < 0.6$ & 95 & 6 \\
            $0.2\leq \text{Sim} < 0.4$ & 170 & 10 \\
            $0\leq \text{Sim} < 0.2$ & 618 & 38 \\
            \bottomrule
        \end{tabular}
    }
    \caption{Comparison between testing result and semantic similarity for Ansible. The overall accuracy of semantic similarity with respect to testing is 84.21\%.}
    \label{tab:ansible-result}
\end{table}

\section{Additional Training Details}
\label{appendix:training}

\textbf{Dataset Split.} The sizes of train, validation, and test datasets for each language are shown in Table \ref{tab:dataset-split}.

\begin{table}[!htb]
    \centering
    \small
    \begin{tabular}{l|ccc|}
        \toprule
        & Train & Validation & Test \\
        \midrule
        Ansible & 393 & 98 & 596 \\
        Bash & 349 & 118 & 585 \\
        SQL & 664 & 177 & 1034 \\
        Lean & 781 & 195 & 520 \\
        \bottomrule
    \end{tabular}
    \caption{Dataset split for each language. }
    \label{tab:dataset-split}
\end{table}

\textbf{Leakage Prevention.} For each language, we manually collected 20 examples from the dataset that cover most functions and utilities of the language. For NL2Bash and Spider dataset, since the train-test splits are already provided, we only selected the examples from the training set. For Ansible dataset and NuminaMath, we removed the selected examples from the dataset to prevent any leakage. We then used the selected examples to synthesize repair data with LLMs as the training dataset.

\textbf{RL Feedback Mechanism.} For token-level updates, we followed the practice outlined by the official documentation of verl to assign the reward only to the last token in the input sequence. For sample trajectory collection, we sampled one program for each input with the current policy, which is passed into syntactic and semantic checkers to calculate the reward.

\section{Cases of Semantic Degradation in Ansible}

\begin{tcolorbox}[breakable,title={Example 1}]
Prompt: Generate an Ansible playbook that sets up a test environment by modifying the hosts file, updating APT on Debian systems, and ensuring wget is installed.
\tcblower
Ground-truth playbook: 
\begin{lstlisting}
---
- name: Prepare the test environment
  hosts: all
  tasks:
  - name: Add a test line to the hosts file to ensure it isn't overwritten
    become: true
    lineinfile:
      path: /etc/hosts_TEST
      line: '# File not overwritten'
      state: present
      insertbefore: BOF
      create: true
      owner: root
      group: root
      mode: '0644'

  - name: Ensure APT is up to date (Debian only)
    apt:
      update_cache: true
    when: ansible_os_family == 'Debian'

  - name: Ensure wget is installed
    package:
      name: wget
      state: present
\end{lstlisting}

Original playbook: 
\begin{lstlisting}
---
- name: Set up test environment
  hosts: all
  become: true
  gather_facts: true

  tasks:
    - name: Ensure test entry is present in hosts file
      ansible.builtin.lineinfile:
        path: /etc/hosts
        line: "127.0.0.1 test.local"
        state: present
        create: false
        insertafter: EOF

    - name: Update APT package cache on Debian systems
      ansible.builtin.apt:
        update_cache: true
      when: ansible_facts.os_family == "Debian"

    - name: Ensure wget is installed
      ansible.builtin.package:
        name: wget
        state: present
\end{lstlisting}

Updated playbook: 
\begin{lstlisting}
---
- name: Set up test environment
  hosts: all
  become: true
  tasks:
  - name: Ensure test entry is present in hosts file
    lineinfile:
      path: /etc/hosts
      line: "127.0.0.1 test.local"
      state: present
      create: false
\end{lstlisting}
\end{tcolorbox}

\begin{tcolorbox}[breakable,title={Example 2}]
Prompt: Generate an Ansible playbook that creates a user group and a user, and sets up a private bin directory for the user.
\tcblower
Ground-truth playbook: 
\begin{lstlisting}
---
- name: Prepare
  hosts: node1
  become: true
  tasks:
  - name: Ensure a user group exists
    ansible.builtin.group:
      name: user
      state: present

  - name: Ensure a normal user exists
    ansible.builtin.user:
      name: k3suser
      group: user
      state: present

  - name: Ensure a normal user has bin directory
    ansible.builtin.file:
      path: /home/k3suser/bin
      state: directory
      owner: k3suser
      group: user
      mode: '0700'
\end{lstlisting}

Original playbook: 
\begin{lstlisting}
- name: Create group, user, and private bin directory
  hosts: all
  become: true
  vars:
    group_name: mygroup
    user_name: myuser
    user_home: "/home/{{ user_name }}"
    private_bin_dir: "{{ user_home }}/bin"

  tasks:
    - name: Create the group
      ansible.builtin.group:
        name: "{{ group_name }}"
        state: present

    - name: Create the user
      ansible.builtin.user:
        name: "{{ user_name }}"
        group: "{{ group_name }}"
        create_home: true
        state: present

    - name: Ensure private bin directory exists
      ansible.builtin.file:
        path: "{{ private_bin_dir }}"
        state: directory
        owner: "{{ user_name }}"
        group: "{{ group_name }}"
        mode: "0750"

    - name: Ensure user's home directory permissions are set
      ansible.builtin.file:
        path: "{{ user_home }}"
        state: directory
        owner: "{{ user_name }}"
        group: "{{ group_name }}"
        mode: "0750"
\end{lstlisting}

Updated playbook: 
\begin{lstlisting}
- name: Create group, user, and private bin directory
  hosts: all
  become: true
  tasks:
  - name: Create the group
    lineinfile:
      path: /etc/group
      regexp: '^mygroup$'
      line: 'mygroup: myuser'
\end{lstlisting}
\end{tcolorbox}

\begin{tcolorbox}[breakable,title={Example 3}]
Prompt: Generate an Ansible playbook that sets up and configures Docker on the localhost.
\tcblower
Ground-truth playbook: 
\begin{lstlisting}
---
- hosts: localhost
  remote_user: root
  tasks:
  - name: Installing Docker
    package:
      name: docker
      state: present
  - name: Starting and Enabling Docker services
    service:
      name: docker
      state: started
      enabled: true
\end{lstlisting}

Original playbook: 
\begin{lstlisting}
---
- name: Set up and configure Docker on localhost
  hosts: localhost
  become: true
  gather_facts: true

  vars:
    docker_packages:
      - docker.io
      - python3-docker
    docker_service_name: docker

  tasks:
    - name: Ensure Docker packages are installed
      ansible.builtin.package:
        name: "{{ docker_packages }}"
        state: present

    - name: Ensure Docker service is enabled and started
      ansible.builtin.service:
        name: "{{ docker_service_name }}"
        enabled: true
        state: started
\end{lstlisting}

Updated playbook: 
\begin{lstlisting}
---
- name: Set up and configure Docker on localhost
  hosts: localhost
  become: true
  tasks:
  - name: Install Docker
    apt:
      name: docker.io
      state: present
\end{lstlisting}
\end{tcolorbox}

\section{Prompt Templates}

\subsection{Code Generation Prompts}

\begin{tcolorbox}[title={Ansible Generation Prompt}]
\begin{lstlisting}
You are an expert in Ansible. The user will give you a task description and ask you to generate an Ansible playbook to complete the given task. You only need to output the content of the playbook. DO NOT use any shell commands (ansible.builtin.shell, ansible.builtin.command, etc.) in the playbook.
Task: {task}
Answer: ```yaml
\end{lstlisting}
\end{tcolorbox}

\begin{tcolorbox}[title={Bash Generation Prompt}]
\begin{lstlisting}
You are an expert in Bash. The user will give you a task description and ask you to generate a bash command to complete the given task. You only need to output the content of the command.
Task: {task}
Answer: ```bash
\end{lstlisting}
\end{tcolorbox}

\begin{tcolorbox}[title={SQL Generation Prompt}]
\begin{lstlisting}
You are an expert in SQL. The user will give you a task description and ask you to generate a SQL command to complete the given task. You only need to output the content of the command.
Task: {task}
Answer: ```sql
\end{lstlisting}
\end{tcolorbox}

\begin{tcolorbox}[title={Lean Generation Prompt}]
\begin{lstlisting}
You are an expert in Lean 4 theorem proving and functional programming. The user will give you a task description and ask you to generate Lean 4 code (definitions, theorems, and proofs) to complete the given task. You only need to output the Lean 4 code. Do NOT use `sorry` in your proofs.
Task: {task}
Answer: ```lean
\end{lstlisting}
\end{tcolorbox}

\subsection{Program Repair Prompts}

\begin{tcolorbox}[title={Ansible Program Repair Prompt}]
\begin{lstlisting}
You are an expert in Ansible. You are asked to fix a possibly incorrect Ansible playbook. You will be provided with the playbook to fix, the user input, and feedback from an interpreter that lists all syntactic errors in the playbook. Your goal is to fix the syntactic errors in the playbook (if any) while following the user's instruction. You only need to output the content of the modified playbook. 

User query: {query}

Original playbook: 
{output}

Interpreter feedback: 
{feedback}

Answer: ```yaml
\end{lstlisting}
\end{tcolorbox}

\begin{tcolorbox}[title={Bash Program Repair Prompt}]
\begin{lstlisting}
You are an expert in Bash. You are asked to fix a possibly incorrect Bash command. You will be provided with the command to fix, the user input, and feedback from an interpreter that lists all syntactic errors in the command. Your goal is to fix the syntactic errors in the command (if any) while following the user's instruction. You only need to output the content of the modified command. 

User query: {query}

Original command: 
{output}

Interpreter feedback: 
{feedback}

Answer: ```bash
\end{lstlisting}
\end{tcolorbox}

\begin{tcolorbox}[title={SQL Program Repair Prompt}]
\begin{lstlisting}
You are an expert in SQL. You are asked to fix a possibly incorrect SQL command. You will be provided with the command to fix, the user input, and feedback from an interpreter that lists all syntactic errors in the command. Your goal is to fix the syntactic errors in the command (if any) while following the user's instruction. You only need to output the content of the modified command. 
User query: {query}

Original command: 
{output}

Interpreter feedback: 
{feedback}

Answer: ```sql
\end{lstlisting}
\end{tcolorbox}

\begin{tcolorbox}[title={Lean Program Repair Prompt}]
\begin{lstlisting}
You are an expert in Lean 4 theorem proving and functional programming. You are asked to fix possibly incorrect Lean 4 code. You will be provided with the code to fix, the user input, and feedback from the Lean type checker that lists all errors. Your goal is to fix the errors (if any) while following the user's instruction. Do NOT use `sorry`. You only need to output the corrected Lean 4 code.

User query: {query}

Original code: 
{output}

Lean checker feedback: 
{feedback}

Answer: ```lean
\end{lstlisting}
\end{tcolorbox}

\subsection{In-context Learning Prompts}

\begin{tcolorbox}[title={Ansible In-context Learning Prompt}]
\begin{lstlisting}
You are an expert in Ansible. The user will give you a task description and ask you to generate an Ansible playbook to complete the given task. You only need to output the content of the playbook. DO NOT use any shell commands (ansible.builtin.shell, ansible.builtin.command, etc.) in the playbook.
The following are some example input queries and corresponding Ansible playbooks for your reference:
{examples}
Task: {task}
Answer: ```yaml
\end{lstlisting}
\end{tcolorbox}

\begin{tcolorbox}[title={Bash In-context Learning Prompt}]
\begin{lstlisting}
You are an expert in Bash. The user will give you a task description and ask you to generate a bash command to complete the given task. You only need to output the content of the command.
The following are some example input queries and corresponding Bash commands for your reference:
{examples}
Task: {task}
Answer: ```bash
\end{lstlisting}
\end{tcolorbox}

\begin{tcolorbox}[title={SQL In-context Learning Prompt}]
\begin{lstlisting}
You are an expert in SQL. The user will give you a task description and ask you to generate a SQL command to complete the given task. You only need to output the content of the command.
The following are some example input queries and corresponding SQL commands for your reference:
{examples}
Task: {task}
Answer: ```sql
\end{lstlisting}
\end{tcolorbox}

\begin{tcolorbox}[title={Lean In-context Learning Prompt}]
\begin{lstlisting}
You are an expert in Lean 4 theorem proving and functional programming. The user will give you a task description and ask you to generate Lean 4 code (definitions, theorems, and proofs) to complete the given task. Do NOT use `sorry` in your proofs.
The following are some example input queries and corresponding Lean 4 code for your reference:
{examples}
Task: {task}
Answer: ```lean
\end{lstlisting}
\end{tcolorbox}

\subsection{Ansible Dataset Query Generation Prompt}

\begin{tcolorbox}[title={Natural Language Query Generation Prompt}]
\begin{lstlisting}
You are an expert in Ansible. You are asked to write a user prompt for the given Ansible playbook that can be used to generate the playbook. Instead of explicitly describing the functionality of the playbook, the prompt should tell what the user wants to accomplish through the playbook. Write the prompt as short as you can, and start the prompt with: Generate an Ansible playbook that ...
\end{lstlisting}
\end{tcolorbox}

%% file: references.bib
@inproceedings{yu_spider_2018,
	title = {Spider: {A} {Large}-{Scale} {Human}-{Labeled} {Dataset} for {Complex} and {Cross}-{Domain} {Semantic} {Parsing} and {Text}-to-{SQL} {Task}},
	url = {https://doi.org/10.18653/v1/d18-1425},
	doi = {10.18653/V1/D18-1425},
	booktitle = {Proceedings of the 2018 {Conference} on {Empirical} {Methods} in {Natural} {Language} {Processing}, {Brussels}, {Belgium}, {October} 31 - {November} 4, 2018},
	publisher = {Association for Computational Linguistics},
	author = {Yu, Tao and Zhang, Rui and Yang, Kai and Yasunaga, Michihiro and Wang, Dongxu and Li, Zifan and Ma, James and Li, Irene and Yao, Qingning and Roman, Shanelle and Zhang, Zilin and Radev, Dragomir R.},
	editor = {Riloff, Ellen and Chiang, David and Hockenmaier, Julia and Tsujii, Jun'ichi},
	year = {2018},
	pages = {3911--3921},
}

@inproceedings{lin_nl2bash_2018,
	title = {{NL2Bash}: {A} {Corpus} and {Semantic} {Parser} for {Natural} {Language} {Interface} to the {Linux} {Operating} {System}},
	url = {http://www.lrec-conf.org/proceedings/lrec2018/summaries/1021.html},
	booktitle = {Proceedings of the {Eleventh} {International} {Conference} on {Language} {Resources} and {Evaluation}, {LREC} 2018, {Miyazaki}, {Japan}, {May} 7-12, 2018},
	publisher = {European Language Resources Association (ELRA)},
	author = {Lin, Xi Victoria and Wang, Chenglong and Zettlemoyer, Luke and Ernst, Michael D.},
	editor = {Calzolari, Nicoletta and Choukri, Khalid and Cieri, Christopher and Declerck, Thierry and Goggi, Sara and Hasida, Kôiti and Isahara, Hitoshi and Maegaard, Bente and Mariani, Joseph and Mazo, Hélène and Moreno, Asunción and Odijk, Jan and Piperidis, Stelios and Tokunaga, Takenobu},
	year = {2018},
}

@misc{openai_gpt-4_2023,
	title={GPT-4 Technical Report}, 
    author={OpenAI and Josh Achiam and Steven Adler and Sandhini Agarwal and Lama Ahmad and Ilge Akkaya and Florencia Leoni Aleman and Diogo Almeida and Janko Altenschmidt and Sam Altman and Shyamal Anadkat and Red Avila and Igor Babuschkin and Suchir Balaji and Valerie Balcom and Paul Baltescu and Haiming Bao and Mohammad Bavarian and Jeff Belgum and Irwan Bello and Jake Berdine and Gabriel Bernadett-Shapiro and Christopher Berner and Lenny Bogdonoff and Oleg Boiko and Madelaine Boyd and Anna-Luisa Brakman and Greg Brockman and Tim Brooks and Miles Brundage and Kevin Button and Trevor Cai and Rosie Campbell and Andrew Cann and Brittany Carey and Chelsea Carlson and Rory Carmichael and Brooke Chan and Che Chang and Fotis Chantzis and Derek Chen and Sully Chen and Ruby Chen and Jason Chen and Mark Chen and Ben Chess and Chester Cho and Casey Chu and Hyung Won Chung and Dave Cummings and Jeremiah Currier and Yunxing Dai and Cory Decareaux and Thomas Degry and Noah Deutsch and Damien Deville and Arka Dhar and David Dohan and Steve Dowling and Sheila Dunning and Adrien Ecoffet and Atty Eleti and Tyna Eloundou and David Farhi and Liam Fedus and Niko Felix and Simón Posada Fishman and Juston Forte and Isabella Fulford and Leo Gao and Elie Georges and Christian Gibson and Vik Goel and Tarun Gogineni and Gabriel Goh and Rapha Gontijo-Lopes and Jonathan Gordon and Morgan Grafstein and Scott Gray and Ryan Greene and Joshua Gross and Shixiang Shane Gu and Yufei Guo and Chris Hallacy and Jesse Han and Jeff Harris and Yuchen He and Mike Heaton and Johannes Heidecke and Chris Hesse and Alan Hickey and Wade Hickey and Peter Hoeschele and Brandon Houghton and Kenny Hsu and Shengli Hu and Xin Hu and Joost Huizinga and Shantanu Jain and Shawn Jain and Joanne Jang and Angela Jiang and Roger Jiang and Haozhun Jin and Denny Jin and Shino Jomoto and Billie Jonn and Heewoo Jun and Tomer Kaftan and Łukasz Kaiser and Ali Kamali and Ingmar Kanitscheider and Nitish Shirish Keskar and Tabarak Khan and Logan Kilpatrick and Jong Wook Kim and Christina Kim and Yongjik Kim and Jan Hendrik Kirchner and Jamie Kiros and Matt Knight and Daniel Kokotajlo and Łukasz Kondraciuk and Andrew Kondrich and Aris Konstantinidis and Kyle Kosic and Gretchen Krueger and Vishal Kuo and Michael Lampe and Ikai Lan and Teddy Lee and Jan Leike and Jade Leung and Daniel Levy and Chak Ming Li and Rachel Lim and Molly Lin and Stephanie Lin and Mateusz Litwin and Theresa Lopez and Ryan Lowe and Patricia Lue and Anna Makanju and Kim Malfacini and Sam Manning and Todor Markov and Yaniv Markovski and Bianca Martin and Katie Mayer and Andrew Mayne and Bob McGrew and Scott Mayer McKinney and Christine McLeavey and Paul McMillan and Jake McNeil and David Medina and Aalok Mehta and Jacob Menick and Luke Metz and Andrey Mishchenko and Pamela Mishkin and Vinnie Monaco and Evan Morikawa and Daniel Mossing and Tong Mu and Mira Murati and Oleg Murk and David Mély and Ashvin Nair and Reiichiro Nakano and Rajeev Nayak and Arvind Neelakantan and Richard Ngo and Hyeonwoo Noh and Long Ouyang and Cullen O'Keefe and Jakub Pachocki and Alex Paino and Joe Palermo and Ashley Pantuliano and Giambattista Parascandolo and Joel Parish and Emy Parparita and Alex Passos and Mikhail Pavlov and Andrew Peng and Adam Perelman and Filipe de Avila Belbute Peres and Michael Petrov and Henrique Ponde de Oliveira Pinto and Michael and Pokorny and Michelle Pokrass and Vitchyr H. Pong and Tolly Powell and Alethea Power and Boris Power and Elizabeth Proehl and Raul Puri and Alec Radford and Jack Rae and Aditya Ramesh and Cameron Raymond and Francis Real and Kendra Rimbach and Carl Ross and Bob Rotsted and Henri Roussez and Nick Ryder and Mario Saltarelli and Ted Sanders and Shibani Santurkar and Girish Sastry and Heather Schmidt and David Schnurr and John Schulman and Daniel Selsam and Kyla Sheppard and Toki Sherbakov and Jessica Shieh and Sarah Shoker and Pranav Shyam and Szymon Sidor and Eric Sigler and Maddie Simens and Jordan Sitkin and Katarina Slama and Ian Sohl and Benjamin Sokolowsky and Yang Song and Natalie Staudacher and Felipe Petroski Such and Natalie Summers and Ilya Sutskever and Jie Tang and Nikolas Tezak and Madeleine B. Thompson and Phil Tillet and Amin Tootoonchian and Elizabeth Tseng and Preston Tuggle and Nick Turley and Jerry Tworek and Juan Felipe Cerón Uribe and Andrea Vallone and Arun Vijayvergiya and Chelsea Voss and Carroll Wainwright and Justin Jay Wang and Alvin Wang and Ben Wang and Jonathan Ward and Jason Wei and CJ Weinmann and Akila Welihinda and Peter Welinder and Jiayi Weng and Lilian Weng and Matt Wiethoff and Dave Willner and Clemens Winter and Samuel Wolrich and Hannah Wong and Lauren Workman and Sherwin Wu and Jeff Wu and Michael Wu and Kai Xiao and Tao Xu and Sarah Yoo and Kevin Yu and Qiming Yuan and Wojciech Zaremba and Rowan Zellers and Chong Zhang and Marvin Zhang and Shengjia Zhao and Tianhao Zheng and Juntang Zhuang and William Zhuk and Barret Zoph},
    year={2024},
    eprint={2303.08774},
    archivePrefix={arXiv},
    primaryClass={cs.CL},
    url={https://arxiv.org/abs/2303.08774}, 
}

@misc{dubey_llama3_2024,
	title={The Llama 3 Herd of Models}, 
    author={LlamaTeam and AI@Meta},
    year={2024},
    eprint={2407.21783},
    archivePrefix={arXiv},
    primaryClass={cs.AI},
    url={https://arxiv.org/abs/2407.21783}, 
}

@article{openai_gpt-5_2025,
	title = {{OpenAI} {GPT}-5 {System} {Card}},
	volume = {abs/2601.03267},
	url = {https://doi.org/10.48550/arXiv.2601.03267},
	doi = {10.48550/ARXIV.2601.03267},
	journal = {CoRR},
	author = {{OpenAI}},
	year = {2026},
	note = {arXiv: 2601.03267},
}

@article{hui_qwen25-coder_2024,
	title = {Qwen2.5-{Coder} {Technical} {Report}},
	volume = {abs/2409.12186},
	url = {https://doi.org/10.48550/arXiv.2409.12186},
	doi = {10.48550/ARXIV.2409.12186},
	journal = {CoRR},
	author = {Hui, Binyuan and Yang, Jian and Cui, Zeyu and Yang, Jiaxi and Liu, Dayiheng and Zhang, Lei and Liu, Tianyu and Zhang, Jiajun and Yu, Bowen and Dang, Kai and Yang, An and Men, Rui and Huang, Fei and Ren, Xingzhang and Ren, Xuancheng and Zhou, Jingren and Lin, Junyang},
	year = {2024},
	note = {arXiv: 2409.12186},
}

@article{mishra_granite_2024,
	title = {Granite {Code} {Models}: {A} {Family} of {Open} {Foundation} {Models} for {Code} {Intelligence}},
	volume = {abs/2405.04324},
	url = {https://doi.org/10.48550/arXiv.2405.04324},
	doi = {10.48550/ARXIV.2405.04324},
	journal = {CoRR},
	author = {Mishra, Mayank and Stallone, Matt and Zhang, Gaoyuan and Shen, Yikang and Prasad, Aditya and Soria, Adriana Meza and Merler, Michele and Selvam, Parameswaran and Surendran, Saptha and Singh, Shivdeep and Sethi, Manish and Dang, Xuan-Hong and Li, Pengyuan and Wu, Kun-Lung and Zawad, Syed and Coleman, Andrew and White, Matthew and Lewis, Mark and Pavuluri, Raju and Koyfman, Yan and Lublinsky, Boris and Bayser, Maximilien de and Abdelaziz, Ibrahim and Basu, Kinjal and Agarwal, Mayank and Zhou, Yi and Johnson, Chris and Goyal, Aanchal and Patel, Hima and Shah, S. Yousaf and Zerfos, Petros and Ludwig, Heiko and Munawar, Asim and Crouse, Maxwell and Kapanipathi, Pavan and Salaria, Shweta and Calio, Bob and Wen, Sophia and Seelam, Seetharami and Belgodere, Brian and Fonseca, Carlos A. and Singhee, Amith and Desai, Nirmit and Cox, David D. and Puri, Ruchir and Panda, Rameswar},
	year = {2024},
	note = {arXiv: 2405.04324},
}

@article{guo_deepseek-coder_2024,
	title = {{DeepSeek}-{Coder}: {When} the {Large} {Language} {Model} {Meets} {Programming} - {The} {Rise} of {Code} {Intelligence}},
	volume = {abs/2401.14196},
	url = {https://doi.org/10.48550/arXiv.2401.14196},
	doi = {10.48550/ARXIV.2401.14196},
	journal = {CoRR},
	author = {Guo, Daya and Zhu, Qihao and Yang, Dejian and Xie, Zhenda and Dong, Kai and Zhang, Wentao and Chen, Guanting and Bi, Xiao and Wu, Y. and Li, Y. K. and Luo, Fuli and Xiong, Yingfei and Liang, Wenfeng},
	year = {2024},
	note = {arXiv: 2401.14196},
}

@inproceedings{kwon_vllm_2023,
	title = {Efficient {Memory} {Management} for {Large} {Language} {Model} {Serving} with {PagedAttention}},
	booktitle = {Proceedings of the {ACM} {SIGOPS} 29th {Symposium} on {Operating} {Systems} {Principles}},
	author = {Kwon, Woosuk and Li, Zhuohan and Zhuang, Siyuan and Sheng, Ying and Zheng, Lianmin and Yu, Cody Hao and Gonzalez, Joseph E. and Zhang, Hao and Stoica, Ion},
	year = {2023},
}

@misc{anthropic_claude_nodate,
	title = {The {Claude} 3 {Model} {Family}: {Opus}, {Sonnet}, {Haiku}},
	url = {https://www-cdn.anthropic.com/de8ba9b01c9ab7cbabf5c33b80b7bbc618857627/Model_Card_Claude_3.pdf},
	author = {{Anthropic}},
    year = {2024},
}

@article{shi_code_2024,
	title = {From {Code} to {Correctness}: {Closing} the {Last} {Mile} of {Code} {Generation} with {Hierarchical} {Debugging}},
	volume = {abs/2410.01215},
	url = {https://doi.org/10.48550/arXiv.2410.01215},
	doi = {10.48550/ARXIV.2410.01215},
	journal = {CoRR},
	author = {Shi, Yuling and Wang, Songsong and Wan, Chengcheng and Gu, Xiaodong},
	year = {2024},
	note = {arXiv: 2410.01215},
}

@misc{chen_evaluating_2021,
      title={Evaluating Large Language Models Trained on Code}, 
      author={Mark Chen and Jerry Tworek and Heewoo Jun and Qiming Yuan and Henrique Ponde de Oliveira Pinto and Jared Kaplan and Harri Edwards and Yuri Burda and Nicholas Joseph and Greg Brockman and Alex Ray and Raul Puri and Gretchen Krueger and Michael Petrov and Heidy Khlaaf and Girish Sastry and Pamela Mishkin and Brooke Chan and Scott Gray and Nick Ryder and Mikhail Pavlov and Alethea Power and Lukasz Kaiser and Mohammad Bavarian and Clemens Winter and Philippe Tillet and Felipe Petroski Such and Dave Cummings and Matthias Plappert and Fotios Chantzis and Elizabeth Barnes and Ariel Herbert-Voss and William Hebgen Guss and Alex Nichol and Alex Paino and Nikolas Tezak and Jie Tang and Igor Babuschkin and Suchir Balaji and Shantanu Jain and William Saunders and Christopher Hesse and Andrew N. Carr and Jan Leike and Josh Achiam and Vedant Misra and Evan Morikawa and Alec Radford and Matthew Knight and Miles Brundage and Mira Murati and Katie Mayer and Peter Welinder and Bob McGrew and Dario Amodei and Sam McCandlish and Ilya Sutskever and Wojciech Zaremba},
      year={2021},
      eprint={2107.03374},
      archivePrefix={arXiv},
      primaryClass={cs.LG},
      url={https://arxiv.org/abs/2107.03374},
}

@article{li_competition-level_2022,
	title = {Competition-{Level} {Code} {Generation} with {AlphaCode}},
	volume = {abs/2203.07814},
	url = {https://doi.org/10.48550/arXiv.2203.07814},
	doi = {10.48550/ARXIV.2203.07814},
	journal = {CoRR},
	author = {Li, Yujia and Choi, David H. and Chung, Junyoung and Kushman, Nate and Schrittwieser, Julian and Leblond, Rémi and Eccles, Tom and Keeling, James and Gimeno, Felix and Lago, Agustin Dal and Hubert, Thomas and Choy, Peter and d'Autume, Cyprien de Masson and Babuschkin, Igor and Chen, Xinyun and Huang, Po-Sen and Welbl, Johannes and Gowal, Sven and Cherepanov, Alexey and Molloy, James and Mankowitz, Daniel J. and Robson, Esme Sutherland and Kohli, Pushmeet and Freitas, Nando de and Kavukcuoglu, Koray and Vinyals, Oriol},
	year = {2022},
	note = {arXiv: 2203.07814},
}

@inproceedings{zhou_codebertscore_2023,
	title = {{CodeBERTScore}: {Evaluating} {Code} {Generation} with {Pretrained} {Models} of {Code}},
	url = {https://doi.org/10.18653/v1/2023.emnlp-main.859},
	doi = {10.18653/V1/2023.EMNLP-MAIN.859},
	booktitle = {Proceedings of the 2023 {Conference} on {Empirical} {Methods} in {Natural} {Language} {Processing}, {EMNLP} 2023, {Singapore}, {December} 6-10, 2023},
	publisher = {Association for Computational Linguistics},
	author = {Zhou, Shuyan and Alon, Uri and Agarwal, Sumit and Neubig, Graham},
	editor = {Bouamor, Houda and Pino, Juan and Bali, Kalika},
	year = {2023},
	pages = {13921--13937},
}

@article{lozhkov_starcoder_2024,
	title = {{StarCoder} 2 and {The} {Stack} v2: {The} {Next} {Generation}},
	volume = {abs/2402.19173},
	url = {https://doi.org/10.48550/arXiv.2402.19173},
	doi = {10.48550/ARXIV.2402.19173},
	journal = {CoRR},
	author = {Lozhkov, Anton and Li, Raymond and Allal, Loubna Ben and Cassano, Federico and Lamy-Poirier, Joel and Tazi, Nouamane and Tang, Ao and Pykhtar, Dmytro and Liu, Jiawei and Wei, Yuxiang and Liu, Tianyang and Tian, Max and Kocetkov, Denis and Zucker, Arthur and Belkada, Younes and Wang, Zijian and Liu, Qian and Abulkhanov, Dmitry and Paul, Indraneil and Li, Zhuang and Li, Wen-Ding and Risdal, Megan and Li, Jia and Zhu, Jian and Zhuo, Terry Yue and Zheltonozhskii, Evgenii and Dade, Nii Osae Osae and Yu, Wenhao and Krauß, Lucas and Jain, Naman and Su, Yixuan and He, Xuanli and Dey, Manan and Abati, Edoardo and Chai, Yekun and Muennighoff, Niklas and Tang, Xiangru and Oblokulov, Muhtasham and Akiki, Christopher and Marone, Marc and Mou, Chenghao and Mishra, Mayank and Gu, Alex and Hui, Binyuan and Dao, Tri and Zebaze, Armel and Dehaene, Olivier and Patry, Nicolas and Xu, Canwen and McAuley, Julian J. and Hu, Han and Scholak, Torsten and Paquet, Sébastien and Robinson, Jennifer and Anderson, Carolyn Jane and Chapados, Nicolas and al, et},
	year = {2024},
	note = {arXiv: 2402.19173},
}

@misc{el-kishky_competitive_2025,
	   title={Competitive Programming with Large Reasoning Models}, 
      author={OpenAI and Ahmed El-Kishky and Alexander Wei and Andre Saraiva and Borys Minaiev and Daniel Selsam and David Dohan and Francis Song and Hunter Lightman and Ignasi Clavera and Jakub Pachocki and Jerry Tworek and Lorenz Kuhn and Lukasz Kaiser and Mark Chen and Max Schwarzer and Mostafa Rohaninejad and Nat McAleese and o3 contributors and Oleg Mürk and Rhythm Garg and Rui Shu and Szymon Sidor and Vineet Kosaraju and Wenda Zhou},
      year={2025},
      eprint={2502.06807},
      archivePrefix={arXiv},
      primaryClass={cs.LG},
      url={https://arxiv.org/abs/2502.06807},
}

@inproceedings{le_coderl_2022,
	title = {{CodeRL}: {Mastering} {Code} {Generation} through {Pretrained} {Models} and {Deep} {Reinforcement} {Learning}},
	url = {http://papers.nips.cc/paper\_files/paper/2022/hash/8636419dea1aa9fbd25fc4248e702da4-Abstract-Conference.html},
	booktitle = {Advances in {Neural} {Information} {Processing} {Systems} 35: {Annual} {Conference} on {Neural} {Information} {Processing} {Systems} 2022, {NeurIPS} 2022, {New} {Orleans}, {LA}, {USA}, {November} 28 - {December} 9, 2022},
	author = {Le, Hung and Wang, Yue and Gotmare, Akhilesh Deepak and Savarese, Silvio and Hoi, Steven Chu-Hong},
	editor = {Koyejo, Sanmi and Mohamed, S. and Agarwal, A. and Belgrave, Danielle and Cho, K. and Oh, A.},
	year = {2022},
}

@article{liu_rltf_2023,
	title = {{RLTF}: {Reinforcement} {Learning} from {Unit} {Test} {Feedback}},
	volume = {2023},
	url = {https://openreview.net/forum?id=hjYmsV6nXZ},
	journal = {Trans. Mach. Learn. Res.},
	author = {Liu, Jiate and Zhu, Yiqin and Xiao, Kaiwen and Fu, Qiang and Han, Xiao and Yang, Wei and Ye, Deheng},
	year = {2023},
}

@article{shojaee_execution-based_2023,
	title = {Execution-based {Code} {Generation} using {Deep} {Reinforcement} {Learning}},
	volume = {2023},
	url = {https://openreview.net/forum?id=0XBuaxqEcG},
	journal = {Trans. Mach. Learn. Res.},
	author = {Shojaee, Parshin and Jain, Aneesh and Tipirneni, Sindhu and Reddy, Chandan K.},
	year = {2023},
}

@inproceedings{dou_stepcoder_2024,
	title = {{StepCoder}: {Improving} {Code} {Generation} with {Reinforcement} {Learning} from {Compiler} {Feedback}},
	url = {https://doi.org/10.18653/v1/2024.acl-long.251},
	doi = {10.18653/V1/2024.ACL-LONG.251},
	booktitle = {Proceedings of the 62nd {Annual} {Meeting} of the {Association} for {Computational} {Linguistics} ({Volume} 1: {Long} {Papers}), {ACL} 2024, {Bangkok}, {Thailand}, {August} 11-16, 2024},
	publisher = {Association for Computational Linguistics},
	author = {Dou, Shihan and Liu, Yan and Jia, Haoxiang and Zhou, Enyu and Xiong, Limao and Shan, Junjie and Huang, Caishuang and Wang, Xiao and Fan, Xiaoran and Xi, Zhiheng and Zhou, Yuhao and Ji, Tao and Zheng, Rui and Zhang, Qi and Gui, Tao and Huang, Xuanjing},
	editor = {Ku, Lun-Wei and Martins, Andre and Srikumar, Vivek},
	year = {2024},
	pages = {4571--4585},
}

@article{jha_rlsf_2024,
	title = {{RLSF}: {Reinforcement} {Learning} via {Symbolic} {Feedback}},
	volume = {abs/2405.16661},
	url = {https://doi.org/10.48550/arXiv.2405.16661},
	doi = {10.48550/ARXIV.2405.16661},
	journal = {CoRR},
	author = {Jha, Piyush and Jana, Prithwish and Arora, Arnav and Ganesh, Vijay},
	year = {2024},
	note = {arXiv: 2405.16661},
}

@inproceedings{zhang_self-edit_2023,
	title = {Self-{Edit}: {Fault}-{Aware} {Code} {Editor} for {Code} {Generation}},
	url = {https://doi.org/10.18653/v1/2023.acl-long.45},
	doi = {10.18653/V1/2023.ACL-LONG.45},
	booktitle = {Proceedings of the 61st {Annual} {Meeting} of the {Association} for {Computational} {Linguistics} ({Volume} 1: {Long} {Papers}), {ACL} 2023, {Toronto}, {Canada}, {July} 9-14, 2023},
	publisher = {Association for Computational Linguistics},
	author = {Zhang, Kechi and Li, Zhuo and Li, Jia and Li, Ge and Jin, Zhi},
	editor = {Rogers, Anna and Boyd-Graber, Jordan L. and Okazaki, Naoaki},
	year = {2023},
	pages = {769--787},
}

@inproceedings{liu_fastfixer_2024,
	title = {{FastFixer}: {An} {Efficient} and {Effective} {Approach} for {Repairing} {Programming} {Assignments}},
	url = {https://doi.org/10.1145/3691620.3695062},
	doi = {10.1145/3691620.3695062},
	booktitle = {Proceedings of the 39th {IEEE}/{ACM} {International} {Conference} on {Automated} {Software} {Engineering}, {ASE} 2024, {Sacramento}, {CA}, {USA}, {October} 27 - {November} 1, 2024},
	publisher = {ACM},
	author = {Liu, Fang and Liu, Zhenwei and Zhao, Qianhui and Jiang, Jing and Zhang, Li and Sun, Zian and Li, Ge and Li, Zhongqi and Ma, Yuchi},
	editor = {Filkov, Vladimir and Ray, Baishakhi and Zhou, Minghui},
	year = {2024},
	pages = {669--680},
}

@inproceedings{ngassom_chain_2024,
	title = {Chain of {Targeted} {Verification} {Questions} to {Improve} the {Reliability} of {Code} {Generated} by {LLMs}},
	url = {https://doi.org/10.1145/3664646.3664772},
	doi = {10.1145/3664646.3664772},
	booktitle = {Proceedings of the 1st {ACM} {International} {Conference} on {AI}-{Powered} {Software}, {AIware} 2024, {Porto} de {Galinhas}, {Brazil}, {July} 15-16, 2024},
	publisher = {ACM},
	author = {Ngassom, Sylvain Kouemo and Dakhel, Arghavan Moradi and Tambon, Florian and Khomh, Foutse},
	editor = {Adams, Bram and Zimmermann, Thomas and Ozkaya, Ipek and Lin, Dayi and Zhang, Jie M.},
	year = {2024},
}

@inproceedings{wang_compilable_2022,
	title = {Compilable {Neural} {Code} {Generation} with {Compiler} {Feedback}},
	url = {https://doi.org/10.18653/v1/2022.findings-acl.2},
	doi = {10.18653/V1/2022.FINDINGS-ACL.2},
	booktitle = {Findings of the {Association} for {Computational} {Linguistics}: {ACL} 2022, {Dublin}, {Ireland}, {May} 22-27, 2022},
	publisher = {Association for Computational Linguistics},
	author = {Wang, Xin and Wang, Yasheng and Wan, Yao and Mi, Fei and Li, Yitong and Zhou, Pingyi and Liu, Jin and Wu, Hao and Jiang, Xin and Liu, Qun},
	editor = {Muresan, Smaranda and Nakov, Preslav and Villavicencio, Aline},
	year = {2022},
	pages = {9--19},
}

@article{jain_tuning_2023,
	title = {Tuning {Models} of {Code} with {Compiler}-{Generated} {Reinforcement} {Learning} {Feedback}},
	volume = {abs/2305.18341},
	url = {https://doi.org/10.48550/arXiv.2305.18341},
	doi = {10.48550/ARXIV.2305.18341},
	journal = {CoRR},
	author = {Jain, Abhinav and Adiole, Chima and Chaudhuri, Swarat and Reps, Thomas W. and Jermaine, Chris},
	year = {2023},
	note = {arXiv: 2305.18341},
}

@article{ye_process-supervised_2025,
	title = {Process-{Supervised} {Reinforcement} {Learning} for {Code} {Generation}},
	volume = {abs/2502.01715},
	url = {https://doi.org/10.48550/arXiv.2502.01715},
	doi = {10.48550/ARXIV.2502.01715},
	journal = {CoRR},
	author = {Ye, Yufan and Zhang, Ting and Jiang, Wenbin and Huang, Hua},
	year = {2025},
	note = {arXiv: 2502.01715},
}

@inproceedings{zhang_bridge-coder_2025,
	title = {Bridge-{Coder}: {Transferring} {Model} {Capabilities} from {High}-{Resource} to {Low}-{Resource} {Programming} {Language}},
	url = {https://aclanthology.org/2025.findings-acl.567/},
	booktitle = {Findings of the {Association} for {Computational} {Linguistics}, {ACL} 2025, {Vienna}, {Austria}, {July} 27 - {August} 1, 2025},
	publisher = {Association for Computational Linguistics},
	author = {Zhang, Jipeng and Zhang, Jianshu and Li, Yuanzhe and Pi, Renjie and Pan, Rui and Liu, Runtao and Zheng, Ziqiang and Zhang, Tong},
	editor = {Che, Wanxiang and Nabende, Joyce and Shutova, Ekaterina and Pilehvar, Mohammad Taher},
	year = {2025},
	pages = {10865--10882},
}

@article{gong_multicoder_2022,
	title = {{MultiCoder}: {Multi}-{Programming}-{Lingual} {Pre}-{Training} for {Low}-{Resource} {Code} {Completion}},
	volume = {abs/2212.09666},
	url = {https://doi.org/10.48550/arXiv.2212.09666},
	doi = {10.48550/ARXIV.2212.09666},
	journal = {CoRR},
	author = {Gong, Zi and Guo, Yinpeng and Zhou, Pingyi and Gao, Cuiyun and Wang, Yasheng and Xu, Zenglin},
	year = {2022},
	note = {arXiv: 2212.09666},
}

@article{sheng_slm-sql_2025,
	title = {{SLM}-{SQL}: {An} {Exploration} of {Small} {Language} {Models} for {Text}-to-{SQL}},
	volume = {abs/2507.22478},
	url = {https://doi.org/10.48550/arXiv.2507.22478},
	doi = {10.48550/ARXIV.2507.22478},
	journal = {CoRR},
	author = {Sheng, Lei and Xu, Shuai-Shuai},
	year = {2025},
	note = {arXiv: 2507.22478},
}

@article{kusama_how_2025,
	title = {How {Small} is {Enough}? {Empirical} {Evidence} of {Quantized} {Small} {Language} {Models} for {Automated} {Program} {Repair}},
	volume = {abs/2508.16499},
	url = {https://doi.org/10.48550/arXiv.2508.16499},
	doi = {10.48550/ARXIV.2508.16499},
	journal = {CoRR},
	author = {Kusama, Kazuki and Shu, Honglin and Kondo, Masanari and Kamei, Yasutaka},
	year = {2025},
	note = {arXiv: 2508.16499},
}

@article{koutcheme_benchmarking_2024,
	title = {Benchmarking {Educational} {Program} {Repair}},
	volume = {abs/2405.05347},
	url = {https://doi.org/10.48550/arXiv.2405.05347},
	doi = {10.48550/ARXIV.2405.05347},
	journal = {CoRR},
	author = {Koutcheme, Charles and Dainese, Nicola and Sarsa, Sami and Leinonen, Juho and Hellas, Arto and Denny, Paul},
	year = {2024},
	note = {arXiv: 2405.05347},
}

@inproceedings{pujar_invited_2023,
	title = {Invited: {Automated} {Code} generation for {Information} {Technology} {Tasks} in {YAML} through {Large} {Language} {Models}},
	url = {https://doi.org/10.1109/DAC56929.2023.10247987},
	doi = {10.1109/DAC56929.2023.10247987},
	booktitle = {60th {ACM}/{IEEE} {Design} {Automation} {Conference}, {DAC} 2023, {San} {Francisco}, {CA}, {USA}, {July} 9-13, 2023},
	publisher = {IEEE},
	author = {Pujar, Saurabh and Buratti, Luca and Guo, Xiaojie and Dupuis, Nicolas and Lewis, Burn L. and Suneja, Sahil and Sood, Atin and Nalawade, Ganesh and Jones, Matthew and Morari, Alessandro and Puri, Ruchir},
	year = {2023},
	pages = {1--4},
}

@inproceedings{liang_grammar-based_2025,
	title = {Grammar-{Based} {Code} {Representation}: {Is} {It} a {Worthy} {Pursuit} for {LLMs}?},
	url = {https://aclanthology.org/2025.findings-acl.807/},
	booktitle = {Findings of the {Association} for {Computational} {Linguistics}, {ACL} 2025, {Vienna}, {Austria}, {July} 27 - {August} 1, 2025},
	publisher = {Association for Computational Linguistics},
	author = {Liang, Qingyuan and Zhang, Zhao and Sun, Zeyu and Lin, Zheng and Luo, Qi and Xiao, Yueyi and Chen, Yizhou and Zhang, Yuqun and Zhang, Haotian and Zhang, Lu and Chenbin, Chenbin and Xiong, Yingfei},
	editor = {Che, Wanxiang and Nabende, Joyce and Shutova, Ekaterina and Pilehvar, Mohammad Taher},
	year = {2025},
	pages = {15640--15653},
}

@misc{councilman_towards_2025,
	   title={Towards Formal Verification of LLM-Generated Code from Natural Language Prompts}, 
      author={Aaron Councilman and David Fu and Aryan Gupta and Chengxiao Wang and David Grove and Yu-Xiong Wang and Vikram Adve},
      year={2025},
      eprint={2507.13290},
      archivePrefix={arXiv},
      primaryClass={cs.PL},
      url={https://arxiv.org/abs/2507.13290}, 
}

@misc{comanici_gemini_2025,
	   title={Gemini 2.5: Pushing the Frontier with Advanced Reasoning, Multimodality, Long Context, and Next Generation Agentic Capabilities}, 
      author={Gemini Team, Google},
      year={2025},
      eprint={2507.06261},
      archivePrefix={arXiv},
      primaryClass={cs.CL},
      url={https://arxiv.org/abs/2507.06261}, 
}

@inproceedings{song_revisiting_2024,
	title = {Revisiting {Code} {Similarity} {Evaluation} with {Abstract} {Syntax} {Tree} {Edit} {Distance}},
	url = {https://doi.org/10.18653/v1/2024.acl-short.3},
	doi = {10.18653/V1/2024.ACL-SHORT.3},
	booktitle = {Proceedings of the 62nd {Annual} {Meeting} of the {Association} for {Computational} {Linguistics}, {ACL} 2024 - {Short} {Papers}, {Bangkok}, {Thailand}, {August} 11-16, 2024},
	publisher = {Association for Computational Linguistics},
	author = {Song, Yewei and Lothritz, Cedric and Tang, Xunzhu and Bissyandé, Tegawendé F. and Klein, Jacques},
	editor = {Ku, Lun-Wei and Martins, Andre and Srikumar, Vivek},
	year = {2024},
	pages = {38--46},
}

@article{song_empirical_2023,
	title = {An {Empirical} {Study} of {Code} {Generation} {Errors} made by {Large} {Language} {Models}},
	language = {en},
	author = {Song, Da and Zhou, Zijie and Wang, Zhijie},
	year = {2023},
}

@article{zhang_llm_2025,
	title = {{LLM} {Hallucinations} in {Practical} {Code} {Generation}: {Phenomena}, {Mechanism}, and {Mitigation}},
	volume = {2},
	url = {https://doi.org/10.1145/3728894},
	doi = {10.1145/3728894},
	number = {ISSTA},
	journal = {Proc. ACM Softw. Eng.},
	author = {Zhang, Ziyao and Wang, Chong and Wang, Yanlin and Shi, Ensheng and Ma, Yuchi and Zhong, Wanjun and Chen, Jiachi and Mao, Mingzhi and Zheng, Zibin},
	year = {2025},
	pages = {481--503},
}

@inproceedings{orlanski_measuring_2023,
	series = {Proceedings of {Machine} {Learning} {Research}},
	title = {Measuring the {Impact} of {Programming} {Language} {Distribution}},
	volume = {202},
	url = {https://proceedings.mlr.press/v202/orlanski23a.html},
	booktitle = {International {Conference} on {Machine} {Learning}, {ICML} 2023, 23-29 {July} 2023, {Honolulu}, {Hawaii}, {USA}},
	publisher = {PMLR},
	author = {Orlanski, Gabriel and Xiao, Kefan and Garcia, Xavier and Hui, Jeffrey and Howland, Joshua and Malmaud, Jonathan and Austin, Jacob and Singh, Rishabh and Catasta, Michele},
	editor = {Krause, Andreas and Brunskill, Emma and Cho, Kyunghyun and Engelhardt, Barbara and Sabato, Sivan and Scarlett, Jonathan},
	year = {2023},
	pages = {26619--26645},
}

@article{sheng2024hybridflow,
  title   = {HybridFlow: A Flexible and Efficient RLHF Framework},
  author  = {Guangming Sheng and Chi Zhang and Zilingfeng Ye and Xibin Wu and Wang Zhang and Ru Zhang and Yanghua Peng and Haibin Lin and Chuan Wu},
  year    = {2024},
  journal = {arXiv preprint arXiv: 2409.19256}
}

@article{yang_qwen25_2024,
	title = {Qwen2.5 {Technical} {Report}},
	volume = {abs/2412.15115},
	url = {https://doi.org/10.48550/arXiv.2412.15115},
	doi = {10.48550/ARXIV.2412.15115},
	journal = {CoRR},
	author = {Yang, An and Yang, Baosong and Zhang, Beichen and Hui, Binyuan and Zheng, Bo and Yu, Bowen and Li, Chengyuan and Liu, Dayiheng and Huang, Fei and Wei, Haoran and Lin, Huan and Yang, Jian and Tu, Jianhong and Zhang, Jianwei and Yang, Jianxin and Yang, Jiaxi and Zhou, Jingren and Lin, Junyang and Dang, Kai and Lu, Keming and Bao, Keqin and Yang, Kexin and Yu, Le and Li, Mei and Xue, Mingfeng and Zhang, Pei and Zhu, Qin and Men, Rui and Lin, Runji and Li, Tianhao and Xia, Tingyu and Ren, Xingzhang and Ren, Xuancheng and Fan, Yang and Su, Yang and Zhang, Yichang and Wan, Yu and Liu, Yuqiong and Cui, Zeyu and Zhang, Zhenru and Qiu, Zihan},
	year = {2024},
	note = {arXiv: 2412.15115},
}

@InProceedings{ruiqi20,
  author =  {Ruiqi Zhong and Tao Yu and Dan Klein},
  title =   {Semantic Evaluation for Text-to-SQL with Distilled Test Suite},
  year =    {2020},
  booktitle =   {The 2020 Conference on Empirical Methods in Natural Language Processing},
  publisher = {Association for Computational Linguistics},
}

@inproceedings{hu_lora_2022,
	title = {{LoRA}: {Low}-{Rank} {Adaptation} of {Large} {Language} {Models}},
	url = {https://openreview.net/forum?id=nZeVKeeFYf9},
	booktitle = {The {Tenth} {International} {Conference} on {Learning} {Representations}, {ICLR} 2022, {Virtual} {Event}, {April} 25-29, 2022},
	publisher = {OpenReview.net},
	author = {Hu, Edward J. and Shen, Yelong and Wallis, Phillip and Allen-Zhu, Zeyuan and Li, Yuanzhi and Wang, Shean and Wang, Lu and Chen, Weizhu},
	year = {2022},
}

@article{shao_deepseekmath_2024,
	title = {{DeepSeekMath}: {Pushing} the {Limits} of {Mathematical} {Reasoning} in {Open} {Language} {Models}},
	volume = {abs/2402.03300},
	url = {https://doi.org/10.48550/arXiv.2402.03300},
	doi = {10.48550/ARXIV.2402.03300},
	journal = {CoRR},
	author = {Shao, Zhihong and Wang, Peiyi and Zhu, Qihao and Xu, Runxin and Song, Junxiao and Zhang, Mingchuan and Li, Y. K. and Wu, Y. and Guo, Daya},
	year = {2024},
	note = {arXiv: 2402.03300},
}

@inproceedings{waldinger_knowledge_1975,
	title = {Knowledge and {Reasoning} in {Program} {Synthesis}},
	url = {http://ijcai.org/Proceedings/75/Papers/041.pdf},
	booktitle = {Advance {Papers} of the {Fourth} {International} {Joint} {Conference} on {Artificial} {Intelligence}, {Tbilisi}, {Georgia}, {USSR}, {September} 3-8, 1975},
	author = {Waldinger, Richard J. and Manna, Zohar},
	year = {1975},
	pages = {288--295},
}

@incollection{manna_towards_1971,
	series = {Lecture {Notes} in {Mathematics}},
	title = {Towards automatic program synthesis},
	url = {https://doi.org/10.1007/BFb0059702},
	doi = {10.1007/BFB0059702},
	booktitle = {Symposium on {Semantics} of {Algorithmic} {Languages}},
	publisher = {Springer},
	author = {Manna, Zohar and Waldinger, Richard J.},
	editor = {Engeler, Erwin},
	year = {1971},
	pages = {270--310},
}

@inproceedings{manna_deductive_1979,
	title = {A {Deductive} {Approach} to {Program} {Synthesis}},
	booktitle = {Proceedings of the {Sixth} {International} {Joint} {Conference} on {Artificial} {Intelligence}, {IJCAI} 79, {Tokyo}, {Japan}, {August} 20-23, 1979, 2 {Volumes}},
	publisher = {William Kaufmann},
	author = {Manna, Zohar and Waldinger, Richard},
	editor = {Buchanan, Bruce G.},
	year = {1979},
	pages = {542--551},
}

@inproceedings{xia_automated_2023,
	title = {Automated {Program} {Repair} in the {Era} of {Large} {Pre}-trained {Language} {Models}},
	url = {https://doi.org/10.1109/ICSE48619.2023.00129},
	doi = {10.1109/ICSE48619.2023.00129},
	booktitle = {45th {IEEE}/{ACM} {International} {Conference} on {Software} {Engineering}, {ICSE} 2023, {Melbourne}, {Australia}, {May} 14-20, 2023},
	publisher = {IEEE},
	author = {Xia, Chunqiu Steven and Wei, Yuxiang and Zhang, Lingming},
	year = {2023},
	pages = {1482--1494},
}

@misc{granite2024granite,
  title={Granite 3.0 Language Models},
  url={https://github.com/ibm-granite/granite-3.0-language-models/},
  author={Granite Team, IBM},
  month={October},
  year={2024}
}

@article{team_qwen3_2025,
	title = {Qwen3 {Technical} {Report}},
	volume = {abs/2505.09388},
	url = {https://doi.org/10.48550/arXiv.2505.09388},
	doi = {10.48550/ARXIV.2505.09388},
	journal = {CoRR},
	author = {Team, Qwen},
	year = {2025},
	note = {arXiv: 2505.09388},
}

@article{wang_kimina-prover_2025,
	title = {Kimina-{Prover} {Preview}: {Towards} {Large} {Formal} {Reasoning} {Models} with {Reinforcement} {Learning}},
	volume = {abs/2504.11354},
	url = {https://doi.org/10.48550/arXiv.2504.11354},
	doi = {10.48550/ARXIV.2504.11354},
	journal = {CoRR},
	author = {Wang, Haiming and Unsal, Mert and Lin, Xiaohan and Baksys, Mantas and Liu, Junqi and Santos, Marco Dos and Sung, Flood and Vinyes, Marina and Ying, Zhenzhe and Zhu, Zekai and Lu, Jianqiao and Saxcé, Hugues de and Bailey, Bolton and Song, Chendong and Xiao, Chenjun and Zhang, Dehao and Zhang, Ebony and Pu, Frederick and Zhu, Han and Liu, Jiawei and Bayer, Jonas and Michel, Julien and Yu, Longhui and Dreyfus-Schmidt, Léo and Tunstall, Lewis and Pagani, Luigi and Machado, Moreira and Bourigault, Pauline and Wang, Ran and Polu, Stanislas and Barroyer, Thibaut and Li, Wen-Ding and Niu, Yazhe and Fleureau, Yann and Hu, Yangyang and Yu, Zhouliang and Wang, Zihan and Yang, Zhilin and Liu, Zhengying and Li, Jia},
	year = {2025},
	note = {arXiv: 2504.11354},
}

@inproceedings{zhao_repair_2024,
	series = {Findings of {ACL}},
	title = {{RePair}: {Automated} {Program} {Repair} with {Process}-based {Feedback}},
	url = {https://doi.org/10.18653/v1/2024.findings-acl.973},
	doi = {10.18653/V1/2024.FINDINGS-ACL.973},
	booktitle = {Findings of the {Association} for {Computational} {Linguistics}, {ACL} 2024, {Bangkok}, {Thailand} and virtual meeting, {August} 11-16, 2024},
	publisher = {Association for Computational Linguistics},
	author = {Zhao, Yuze and Huang, Zhenya and Ma, Yixiao and Li, Rui and Zhang, Kai and Jiang, Hao and Liu, Qi and Zhu, Linbo and Su, Yu},
	editor = {Ku, Lun-Wei and Martins, Andre and Srikumar, Vivek},
	year = {2024},
	pages = {16415--16429},
}

@inproceedings{gupta_deep_2019,
	title = {Deep {Reinforcement} {Learning} for {Syntactic} {Error} {Repair} in {Student} {Programs}},
	url = {https://doi.org/10.1609/aaai.v33i01.3301930},
	doi = {10.1609/AAAI.V33I01.3301930},
	booktitle = {The {Thirty}-{Third} {AAAI} {Conference} on {Artificial} {Intelligence}, {AAAI} 2019, {The} {Thirty}-{First} {Innovative} {Applications} of {Artificial} {Intelligence} {Conference}, {IAAI} 2019, {The} {Ninth} {AAAI} {Symposium} on {Educational} {Advances} in {Artificial} {Intelligence}, {EAAI} 2019, {Honolulu}, {Hawaii}, {USA}, {January} 27 - {February} 1, 2019},
	publisher = {AAAI Press},
	author = {Gupta, Rahul and Kanade, Aditya and Shevade, Shirish K.},
	year = {2019},
	pages = {930--937},
}
